\documentclass[11 pt]{article}
\usepackage[]{inputenc}
\usepackage{esvect}
\usepackage{amsmath}
\usepackage{amsthm}
\usepackage{comment}
\usepackage{longtable}
\usepackage{cite}
\usepackage{soul}
\usepackage{graphicx}
\usepackage{subcaption}
\usepackage[hidelinks]{hyperref}
\usepackage[normalem]{ulem}
\hypersetup{
    colorlinks,
    linkcolor={red!60!black},
    citecolor={blue!80!black},
    urlcolor={blue!80!black}
}
\usepackage[left=2 cm, right=2.00cm, top=2.5cm, bottom=2.0cm]{geometry}
\usepackage{mmacells}
\usepackage{float}

\newcommand{\mt}{\textsc{Mathematica}\,}
\newcommand{\pd}{Pad\'e \,}
\newcommand{\ca}{\texttt{ChisholmD.wl}\,}
\newcommand{\ft}{\texttt{AppellF2.wl}\,}

\newcommand{\linefill}{
  {-}\mkern-7mu
  \cleaders\hbox{$\mkern-2mu-\mkern-2mu$}\hfill
  \mkern-7mu{-}%
}
\title{\texttt{ChisholmD.wl}- Automated rational approximant for bi-variate series}

\author{Souvik Bera\thanks{\href{mailto:souvikbera@iisc.ac.in}{souvikbera@iisc.ac.in}} \hspace{.1cm} Tanay Pathak\thanks{\href{mailto:tanaypathak@iisc.ac.in}{tanaypathak@iisc.ac.in}}}

\date{Centre for High Energy Physics, Indian Institute of Science,\\ Bangalore-560012, Karnataka, India
}
\begin{document}
\maketitle
\begin{abstract} 
 \noindent The Chisholm rational approximant is a natural generalization to two variables of the well-known single variable \pd approximant, and has the advantage of reducing to the latter when one of the variables is set equals to 0.  We present, to our knowledge, the first automated \textsc{Mathematica} package to evaluate diagonal Chisholm approximants of two variable series. For the moment, the package can only be used to evaluate diagonal approximants i.e. the maximum powers of both the variables, in both the numerator and the denominator, is equal to some integer $M$. We further modify the original method so as to allow us to evaluate the approximants around some general point $(x,y)$ not necessarily $(0,0)$. Using the approximants around general point $(x,y)$, allows us to get a better estimate of the result when the point of evaluation is far from $(0,0)$. Several examples of the elementary functions have been studied which shows that the approximants can be useful for analytic continuation and convergence acceleration purposes. We continue our study using various examples of two variable hypergeometric series, $\mathrm{Li}_{2,2}(x,y)$ etc that arise in particle physics and in the study of critical phenomena in condensed matter physics. The demonstration of the package is discussed in detail and the \textsc{Mathematica} package is provided as an ancillary file.
\end{abstract}

\vspace{2.5cm}

\setlength{\parindent}{20pt}
{\bf Program summary:}
\begin{itemize}
    \item \textit{Program Title}: \ca.
    \vspace{-0.2cm}
    \item \textit{Licensing provisions}: GNU General Public License v3.0.
    \vspace{-0.2cm}
    \item \textit{Programming language}: Wolfram Mathematica version 13.2.0 and beyond.
\vspace{-0.2cm}
    \item \textit{Nature of problem}: To find the diagonal rational approximant of two variable series, analogous to \pd approximant of one variable series.
 \vspace{-0.2cm}
    \item \textit{Solution method}: \mt implementation of Chisholm's method to find the diagonal approximant of the two variable series.
\end{itemize}

\newpage

\section{Introduction}
In practical problems it is possible that only the lower order terms of a series are known \cite{ferris1973numerical, Ananthanarayan:2020umo, AlamKhan:2023dms} and it is desirable to estimate the higher order terms of the series using this information. For a given series in one variable, truncated up to certain order $M$ in its variable, one can estimate the higher order terms by using rational approximants. Rational approximants are approximations of a given truncated series (which is actually infinite series but is known only up to certain order) using a rational function. A popular and widespread rational approximant for series in one variable is the \pd approximants \cite{baker1964theory,george1975essentials,graves1981pade}.
Though in physics applications the two-variable series also appear quite frequently, Appell $F_{1}$ hypergeometric series\cite{appell1926fonctions,olsson1964integration,Srivastava:1985}, $\text{Li}_{2,2}(x,y)$\cite{kummer1840about,poincare1884groups,chen1977iterated,goncharov2001multiple,Goncharov:2010jf} series to name a few. Thus, it is desirable to have approximants analogous to \pd approximants for the multi-variable case. There are various ways to form such multivariate approximations \cite{chisholm1973rational,chisholm1974rational,graves1974calculation,cuyt1999well,cuyt1983multivariate,jones1976general,levin1976general,baker1961pade,cuyt1996nuttall}. Though, in the present work, we are interested in the construction and study of bi-variate rational approximants. The generalisation to the bi-variate case is not straightforward, as the correct number of linear equations to determine the coefficients in the approximation cannot be formed, unlike the case of \pd approximants. Chisholm later proposed a way to obtain the correct number of linear equations \cite{chisholm1973rational}. This bi-variate generalization
of \pd approximant is the Chisholm approximant(CA), which shares some desirable properties similar to the \pd approximant. Apart from other properties, it is also reducible to the case of \pd approximant when one of the variables is set to 0.  

With this motivation, we use the method presented in \cite{chisholm1973rational} to construct approximants of series in two variables. We will also modify the method so as to obtain the CA around any given point $(x,y)$. The method is further implemented in a \mt package \ca(Chisholm Diagonal). We focus only on the construction of the diagonal approximant, which implies that the maximum powers of both the variables, in both the numerator and denominator, is equal to some integer $M$. Though we would like to mention that there are ways to construct off-diagonal CA for both the two variable cases \cite{graves1974calculation,jones1976general}. For the case of one variable the \pd approximant has also been used for the purpose of numerical analytic continuation \cite{ferris1973numerical,baker1996pade}. Motivated by this we further study applications of these approximants for the purpose of analytic continuation and convergence acceleration using examples of multivariable hypergeometric series and generalized multiple polylogarithms(MPLs). It is sometimes possible that some of the coefficients in the general series of a two-variable function are zero and hence the CA approximant does not exist. We discuss how using certain transformations we can obtain the Ca for these series. Some examples of the series with these properties are the ones that arise in the study of critical phenomena in condensed matter physics systems \cite{watson1974two,wood1975applications,wood1974chisholm}. This further shows the application of these approximants in theoretical physics.

The article is organized as follows. We give a brief review of \pd approximants in section \ref{sec:pade} discussing its properties with examples. In section \ref{sec:CA} we will discuss the method to construct the CA given in \cite{chisholm1973rational}. A description of the package \ca is given in section \ref{sec:cades}. We then discuss various examples of elementary functions and applications of these approximants in section \ref{sec:examples} and \ref{sec:application} respectively. This will be followed by a summary and discussion in section \ref{sec:sumdiss}.

\section{Pad\'e approximant}\label{sec:pade}

In this section, we  review \pd approximants \cite{baker1964theory,george1975essentials,graves1981pade}  and discuss some of their application using examples. Pad\'e approximant of a given series is an approximation using a rational function of a given order. Consider a series $f(x)$ around $x=0$
\begin{equation}
    f(x) = \sum_{n=0}^{\infty} a_{n} x^{n}
\end{equation}
The Pad\'e approximant of $f(x)$, denoted by $R_{M/N} \equiv [M/N]$\footnote{When the pade approximant is evaluated around point $x=a$ then we would denote it as $[M/N]_{a}$ and $[M/N]_{0}$ is denoted as $[M/N]$. }, is given by 
\begin{equation} \label{eqn:pade0}
    R_{M/N} =  \frac{\sum_{i=0}^{M} p_{i} x^{i}}{1+\sum_{j=1}^{N} q_{j} x^{j}} = \frac{P(x)}{Q(x)}
\end{equation}
where $P(x)$ and $Q(x)$ are polynomials of degree $M$ and $N$ respectively. When the degree of the polynomial in numerator and denominator is the same then the approximant, $[M/M]$, is called the diagonal \pd approximant.

The coefficients $p_{i}$ and $q_{j}$ can be obtained by setting 
\begin{align}
    f(x) = \frac{P(x)}{Q(x)}
\end{align}
or,
\begin{align}
    \left[1+\sum_{j=1}^{N} q_{j} x^{j}\right] \left[\sum_{n=0}^{\infty} a_{n} x^{n}\right] =  \sum_{i=0}^{M} p_{i} x^{i}
\end{align}
By collecting different powers of $x$, one can find a set of equations, which is needed to be solved. We, hereon, specialise for the case of diagonal approximants i.e. $M=N$, and we get following set of equations
\begin{align}
   p_{0} &= a_{0} \nonumber \\
   p_{1} &= a_{1} + a_{0}q_{1} \nonumber \\
   \vdots \nonumber \\
   p_{M} &= a_{M} + a_{M-1} q_{1} + \cdots + a_{0} q_{M} \nonumber \\
   0 &=  a_{M+1}+ a_{M} q_{1}+ \cdots + a_{1}q_{M} \nonumber  \\ \vdots \nonumber \\
   0 &= a_{2M}+ a_{2M-1}q_{1} + \cdots + a_{M}q_{M}
\end{align}
Further, without loss of generality, one can normalize the series: $p_{0}= a_{0} =1$. With this we have $2M$ unknown coefficients and $2M$ linear equations to solve.  
There exist various algorithms for efficient calculation of these coefficients \cite{wynn1966convergence, brezinski1996extrapolation}. When the solution of the set of linear equations exists, Pad\'e approximant is unique for a given $M$ (and $N$ in general).

\begin{figure}
		\centering
		\begin{subfigure}[b]{0.48\textwidth}
		\centering
		\includegraphics[width=\textwidth]{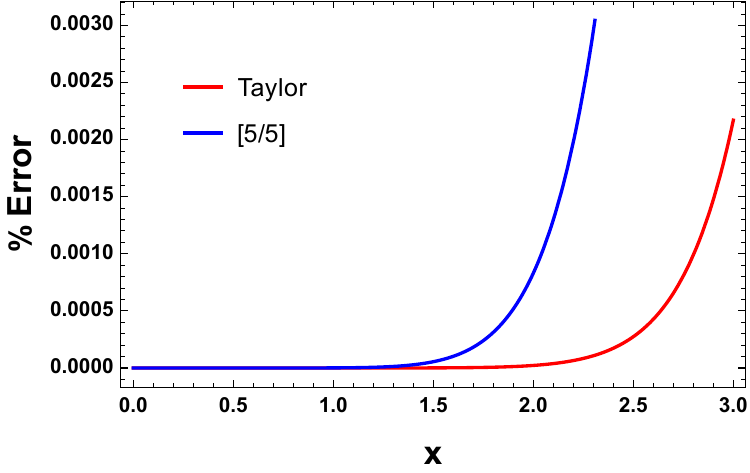}
			\caption{}
			\label{fig:jj1}
		\end{subfigure}
		\hfill
		\begin{subfigure}[b]{0.5\textwidth}
		\centering
		\includegraphics[width=.99\textwidth]{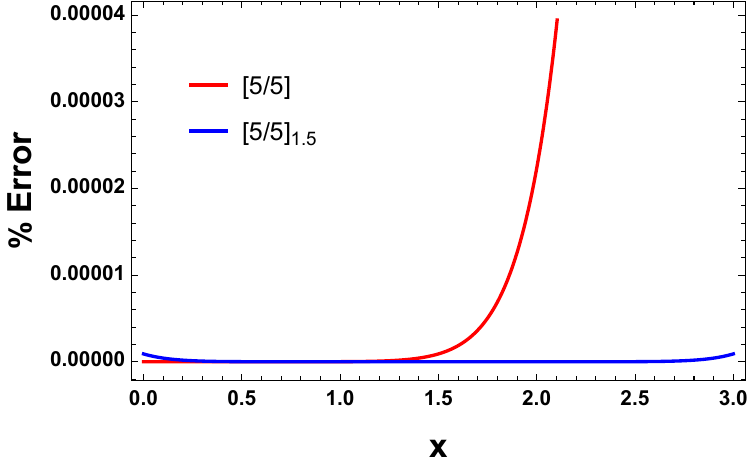}
			\caption{}
			\label{fig:jj2}
		\end{subfigure}
		\caption{(a) Comparison of the percentage error of Taylor series truncated at $\mathcal{O}(x^{10})$ and Pad\'e approximants $[5/5]$ around point $x= 0$, in red and blue respectively. (b) Comparison of the percentage error of Pad\'e approximants $[5/5]$ evaluated around point $x= 0$ and $x= 1.5$, in red and blue respectively. The maximum \% error for the case of $[5/5]_{1.5}$ is of the order of $10^{-7}$ for the range of $x$ shown in the plot. }
		\label{fig:taylor_pade}
\end{figure}

An important feature of the \pd approximant is that it often gives a better approximation of the series than the corresponding truncated Taylor series it is constructed from. To show this, we take the series of $e^{x}$, and compare the result of the  Taylor series truncated at $\mathcal{O}(x^{10})$ with that of \pd approximant $[5/5]$ of $e^{x}$. In Fig.\eqref{fig:jj1} we plot the corresponding percentage error for a range of values of $x$. We clearly see that the results obtained using \pd approximant agree better with the exact function for a larger range of $x$ as compared to the truncated Taylor series.

The \pd approximant obtained using the Eq.\eqref{eqn:pade0} is obtained around the point $x=0$. Due to this the \pd approximant defined using Eq.\eqref{eqn:pade0} tends to deviate from the exact result as we move further away from $x=0$. We can generalize this process and evaluate the \pd approximant around any given point $x= a$ and obtain the approximant $[M/N]_{a}$. This allows us to get better results farther away from $x=0$ and around our point of interest, $x=a$. In Fig.\eqref{fig:jj2} we show the comparison of two \pd approximants obtained around $x=0$ and $x=1.5$ by plotting the percentage error using the two approximants in red and blue respectively. We see from the plot that if our point of interest is $x=2$ then $[5/5]_{1.5}$ agrees better with the exact $e^{x}$, as compared to $[5/5]$.

Another interesting feature of \pd approximant is that it may sometimes provide results outside the domain of convergence of the corresponding Taylor series it is constructed from. As an example, consider the Gauss hypergeometric $_2F_1$ series
\begin{align}
    _2F_1(a,b,c;z) = \sum_{n=0}^\infty \frac{(a)_n (b)_n}{(c)_n} \frac{z^n}{n!}
\end{align}
which is valid in $|z|<1$. Clearly, the points $z= \frac{1}{2} \left(1\pm i \sqrt{3}\right)$ do not belong to the defining domain of convergence of the $_2F_1$ series. It turns out that, the well-known analytic continuations of $_2F_1(\dots; z)$ around $z=1$ or $z=\infty$ can not be used to find the value of the function at these points. A special treatment \cite{buhring1987analytic} is required to evaluate the value of the Gauss $_2F_1$ at these special points.

In the following, we find the value of $_2F_{1}(1/2,1/3,1/5;z)$ at $z= \frac{1}{2} \left(1- i \sqrt{3}\right)$  using the $[10/10]$ \pd approximants and compare with the result obtained using the \mt inbuilt command \texttt{Hypergeometric2F1}.
We first store terms up to $\mathcal{O}(x^{20})$ of the Gauss $_2F_1(1/2,1/3,1/5;z)$ series in the variable \texttt{ser}.
\begin{mmaCell}{Input}
 ser = Normal[Series[Hypergeometric2F1[1/2,1/3,1/5,z],\{z,0,20\}]];
\end{mmaCell}

Next, we find the \pd approximation using the \mt inbuilt command \texttt{PadeApproximant} at $x=0$ of order $10$.

\begin{mmaCell}{Input}
approx = PadeApproximant[ser,\{z,0,10\}];
\end{mmaCell}

Finally, we evaluate the \texttt{approx} at the point $z = \frac{1}{2} \left(1- i \sqrt{3}\right)$ and compare with \mt implementation of \texttt{Hypergeometric2F1}

\begin{mmaCell}{Input}
\{N[approx /.x->\mmaFrac{(1-I\mmaSqrt{3})}{2},10], N[Hypergeometric2F1[\mmaFrac{1}{2},\mmaFrac{1}{3},\mmaFrac{1}{5},\mmaFrac{(1-I\mmaSqrt{3})}{2}],10]\}

\end{mmaCell}
\begin{mmaCell}{Output}
\{0.7062090573-0.8072538749 I,0.7062090573-0.8072538748 I\}
\end{mmaCell}

We see that the result obtained using the \pd approximant is quite accurate and matches well with the implementation in \mt. We also see that we just need 20 terms in the series of $_2F_1$ to obtain the $[10/10]$ approximant. 

\section{Chisholm approximant}\label{sec:CA}

In this section, we  briefly outline the procedure to construct Chisholm approximant (CA) for two variable series  following \cite{chisholm1973rational}.  Consider a bi-variate series of the following form
\begin{equation}\label{eq:genseries}
    f(x,y) 
 = \sum_{m,n=0}^{\infty} c_{mn}x^{m}y^{n}
\end{equation}

We seek rational diagonal approximants of the series above.  Similar to the case of \pd approximant we denote the CA of order $M$ as $[M/M]$. If on the other hand, the CA is obtained around the point $(a,b)$ then it is denoted as $[M/M]_{(a,b)}$. Thus for $f_{M,M}(x,y) \equiv [M/M]$ approximants we have
\begin{equation}
    f_{M,M}(x,y) = \frac{\sum_{p,q=0}^{M} a_{pq}x^{m}y^{n}}{\sum_{r,s=0}^{M} b_{rs}x^{r}y^{s}}
\end{equation}

where $a_{p q}$ and $b_{r s}$ are coefficients to be determined.
Without loss of generality, we can assume that $c_{00}=1$. This will allow us to choose 
\begin{equation*}
    a_{00}= b_{00}=1
\end{equation*}

The total number of coefficients to be determined for $f_{M,M}$ is
\begin{equation}
    2[(M+1)^{2}-1] = 2M^{2}+4M
\end{equation}

Thus, we need the same number of equations to solve the unknown coefficients. To demonstrate how one can construct the required number of equations, we consider the $f_{1,1}$ approximant of the general series given by Eq.\eqref{eq:genseries}.

\begin{equation}\label{eq:Chisholm1}
    f_{1,1}(x,y) = \frac{1+ a_{10}x +a_{01}y+a_{11}x y}{1+ b_{10}x +b_{01}y+b_{11}x y}
\end{equation}

The coefficients of the approximation can be found by setting 
\begin{align}
    f(x,y) = f_{1,1} (x,y)
\end{align}

To get the right number of consistency equations to solve for coefficients we need the following expansion of $f(x,y)$ 
\begin{equation}\label{eq:func11}
    f(x,y) = 1+ c_{10}x + c_{01}y + c_{11}xy + c_{20}x^{2}+c_{02}y^{2}+c_{21}x^{2}y + c_{12}xy^{2} 
\end{equation}
Using Eq. \eqref{eq:Chisholm1} and Eq. \eqref{eq:func11} we now form the following two sets of equations
\begin{itemize}
    \item By comparing coefficients of $x, y, xy, x^{2}$ and $y^{2}$,  we obtain
    \begin{align}\label{eq:consistent1}
        b_{10}+c_{10} &= a_{10} \nonumber \\
        b_{01}+c_{01} &= a_{01} \nonumber \\
        (b_{11}+c_{11})+(b_{10}c_{01} +b_{01}c_{10}) &= a_{11}\nonumber \\
        c_{20}+c_{10}b_{10}&=0 \nonumber \\
        c_{02}+c_{01}b_{01}&=0 \nonumber \\
    \end{align}
    \item From above we see that we have already obtained 5 equations and we have a total of 6 unknowns so we need one more equation. We form it by adding the coefficients of $x^{2}y$ and $xy^{2}$. We thus obtain the following equation
     \begin{align}\label{eq:consistent2}
        (c_{20}+c_{11})b_{01}+(c_{11}+c_{02})b_{10}+ (c_{10}+c_{01})b_{11}+c_{21}+c_{12} &=0   
    \end{align}
\end{itemize}
The above strategy, shown specifically for $[1/1]$ CA is a special case of the general procedure to find  $[M/M]$ CA for any $ M$. The approximants thus obtained, have the following properties 
\begin{itemize}
    \item The approximants are symmetrical in the variables $x$ and $y$. 
    \item The approximants when they exist are unique.
    \item If $x=0$ or $y=0$ then the approximants become the diagonal \pd approximants in the other variable. 
    \item The approximants are invariant under all transformations of the group 
    \begin{equation}
        x = \frac{A u}{1- B u}, \quad y = \frac{A v}{1- B v} \quad (A \neq 0).
    \end{equation}
    \item Consider the reciprocal of the series Eq.  \eqref{eq:genseries} which is given by 
    \begin{equation}\label{eq:genseriesreci}
       \frac{1}{f(x,y)}= \sum_{m,n=0}^{\infty} d_{mn}x^{m}y^{n}
    \end{equation}
    then the reciprocal of the approximant defined from Eq.\eqref{eq:genseries} is equal to the corresponding approximant defined from 
    Eq.\eqref{eq:genseriesreci}.
\end{itemize}
It has been also shown in \cite{chisholm1973rational} that these are the only approximants satisfying all the above properties.

The approximants formed from Eq.\eqref{eq:genseries} are constructed around the point $(x,y) = (0,0)$. Similar to the case of the \pd approximants we can modify the method to obtain the CA around any point $(a,b)$. To do this we need the series of the following form 
\begin{equation}
    \sum_{m,n=0}^{\infty} c'_{mn}(x-a)^{m}(y-b)^{n}
\end{equation}
Analogous to series \eqref{eq:genseries}, we assume the series is of the following form
\begin{equation}\label{eq:genseriesab}
    \sum_{m,n=0}^{\infty} c'_{mn}X^{m}Y^{n}
\end{equation}
where $X=x-a$ and $Y=y-b$.

With the series given in Eq. \eqref{eq:genseriesab}, we now repeat the procedure discussed above and obtain approximant in the new variables $X$ and $Y$. Finally, in the approximant obtained we substitute back $X=x-a$ and $Y=y-b$, to obtain the CA around point $(a,b)$. In a later section \ref{sec:application} we use such a procedure to find CA of series with $X$ and $Y$ as general functions of variables $x$ and $y$.

\section{Description of \ca}\label{sec:cades}
The  method presented in section 
\ref{sec:CA} has been automatized in the accompanying \mt package \ca. We demonstrate the usage of the package \ca below.
After downloading the package and putting it in the same directory as the notebook we can call the package as follows: 
\begin{mmaCell}{Input}
SetDirectory[NotebookDirectory[]];
ChisholmD.wl;
\end{mmaCell}
\begin{mmaCell}{Input}
<<ChisholmD.wl
\end{mmaCell}
\begin{mmaCell}{Print}
ChrisholmD.wl 1.0
Authors : Souvik Bera & Tanay Pathak

\end{mmaCell}

The only command of the package is \texttt{ChisholmD}, which can be called as follows
\begin{mmaCell}{Input}
ChrislholmD[Series,\{x0,y0,order\},\{x,y\}]
\end{mmaCell}

The various elements of the input are given below.
\begin{itemize}
    \item \texttt{Series}: This is the series for which we want to determine its CA. The series is always given around $(0,0)$ even for the cases when the approximant is to be determined around point $(a,b)$.
    \item \texttt{\{x0,y0,order\}}: It is a list containing three elements. \texttt{x0} and \texttt{y0} refer to the point around which the approximant is to be determined and \texttt{order} refers to the required order of the approximant. 
    \item \texttt{\{x,y\}}: It is a list containing two entries. These are the variables of the series and also the resulting approximant.
\end{itemize}
The output of the above command is the CA of the \texttt{Series} of order \texttt{order} around the point (\texttt{x0,y0}).

Let us illustrate the usage of the command by a simple example of double variable series of
 $\exp\left(x+y\right)$. To obtain its CA around $(x,y)=(0,0)$ of order $1$,  we can use the following command, where the series of $\exp\left(x+y\right)$ is stored in the variable \texttt{Expseries}

\begin{mmaCell}{Input}
ChisholmD[Expseries,\{0,0,1\},\{x,y\}]
\end{mmaCell}
\begin{mmaCell}{Output}
\mmaFrac{\mmaFrac{xy}{4}+\mmaFrac{x}{2}+\mmaFrac{y}{2}+1}{\mmaFrac{xy}{4}-\mmaFrac{x}{2}-\mmaFrac{y}{2}+1}

\end{mmaCell}

Note that, the output expression is symmetric in $x$ and $y$. Substituting $y=0$, we obtain
\begin{equation*}
  \frac{1+\frac{x}{2}}{1-\frac{x}{2}}.
\end{equation*}
This expression is the well-known $[1/1]$ \pd approximant of $e^{x}$ and can be easily verified in \mt.
To form the correct number of consistency equations for evaluation of $[M/M]$ approximant(analogous to Eqn.\eqref{eq:consistent1} and \eqref{eq:consistent2}) we need to provide all the terms of the series of the form $x^{\alpha}y^{\beta}$ such that $\alpha+\beta \leq 2M+1, \{ \alpha,\beta\} \neq 2M+1$. Giving extra terms won't affect the computation and the result, but if the correct terms are not present in the series then there would be an error message displayed saying : \texttt{Equations may not give solutions for all solve variables}.

\section{Examples of two-variable series}\label{sec:examples}
In this section, we numerically study the CAs of some elementary series and compare their result with the exact series. 
We will take the examples of elementary functions: $\exp(\frac{x+y}{2}),\, \sin\left(\frac{x+y}{2}\right),\, \sinh\left(\frac{x+y}{2}\right)$ and $\log(1+x+y)$. 

\subsection{\boldmath $\exp\left(\frac{x+y}{2}\right)$}
We can obtain the CA for $\exp(\frac{x+y}{2})$ around $(0,0)$ using the following command
\begin{mmaCell}{Input}
ChisholmD[Normal[Series[Exp[(x+y)/2],\{x,0,20\},\{y,0,20\}]],\{0,0,10\},\{x,y\}]
\end{mmaCell}

The CA of the same function around $(3,6)$ can be obtained in a similar way
\begin{mmaCell}{Input}
ChisholmD[Normal[Series[Exp[(x+y)/2],\{x,0,20\},\{y,0,20\}]],\{3,6,10\},\{x,y\}]
\end{mmaCell}

In Table \eqref{tab:exp00} and \eqref{tab:exp22} we compare the values obtained using the CA and the values obtained using the in-built \mt functions. We also provide the percentage error in the evaluation of values using the CA. The Table \eqref{tab:exp00} corresponds to the values obtained using the CA around $(0,0)$ while the Table \eqref{tab:exp22} corresponds to the values obtained using the CA around $(3,6)$. We observe from Table \eqref{tab:exp00} that the error is less when the chosen points are closer to $(0,0)$ \footnote{A point $(x,y)$ is closer to $(0,0)$ if its Euclidean distance from $(0,0)$ is less as compared to others.}  and worsen as we move away from the $(0,0)$. A similar pattern in the numerical values is observed in the Table \eqref{tab:exp22} also. Thus, for computation purposes if, for example, the point of evaluation is $(5,5)$ then it is better to use $[10/10]_{(3,6)}$ approximant than $[10/10]$ approximant.
\begin{table}[H]
\caption{Table of values of $\exp(\frac{x+y}{2})$}
\begin{subtable}[t]{.5\textwidth}
\raggedright
\begin{tabular}{|c|c|c|c|}
\hline
\{x,y\} & CA          & Function    & \% Error                    \\ \hline
\{0,0\} & 1.000000000 & 1.000000000 & 0                           \\ \hline
\{0,3\} & 4.481689070 & 4.481689070 & $2.2\times10^{-14}$ \\ \hline
\{0,6\} & 20.08553692 & 20.08553692 & $3.2\times10^{-9 }$ \\ \hline
\{0,9\} & 90.01712793 & 90.01713130 & $3.7\times10^{-6 }$ \\ \hline
\{3,0\} & 4.481689070 & 4.481689070 & $2.2\times10^{-14}$ \\ \hline
\{3,3\} & 20.08553692 & 20.08553692 & $4.5\times10^{-14}$ \\ \hline
\{3,6\} & 90.01713130 & 90.01713130 & $3.2\times10^{-9 }$ \\ \hline
\{3,9\} & 403.4287784 & 403.4287935 & $3.7\times10^{-6 }$ \\ \hline
\{6,0\} & 20.08553692 & 20.08553692 & $3.2\times10^{-9}$  \\ \hline
\{6,3\} & 90.01713130 & 90.01713130 & $3.2\times10^{-9 }$ \\ \hline
\{6,6\} & 403.4287935 & 403.4287935 & $6.4\times10^{-9}$  \\ \hline
\{6,9\} & 1808.042347 & 1808.042414 & $3.8\times10^{-6}$  \\ \hline
\{9,0\} & 90.01712793 & 90.01713130 & $3.7\times10^{-6 }$ \\ \hline
\{9,3\} & 403.4287784 & 403.4287935 & $3.7\times10^{-6}$  \\ \hline
\{9,6\} & 1808.042347 & 1808.042414 & $3.8\times10^{-6}$  \\ \hline
\{9,9\} & 8103.083320 & 8103.083928 & $7.5\times10^{-6}$  \\ \hline
\end{tabular}
\vspace{.5cm}
\caption{\label{tab:exp00}Table of values obtained using CA around $(0,0)$.}
\end{subtable}%
\begin{subtable}[t]{.5\textwidth}
\raggedleft
\begin{tabular}{|c|c|c|c|}
\hline
\{x,y\} & CA          & Function    & \% Error                    \\ \hline
\{0,0\} & 1.000000000 & 1.000000000 & $1.2\times10^{-13}$ \\ \hline
\{0,3\} & 4.481689070 & 4.481689070 & $1.1\times10^{-19} $\\ \hline
\{0,6\} & 20.08553692 & 20.08553692 & $5.4\times10^{-20}$ \\ \hline
\{0,9\} & 90.01713130 & 90.01713130 & $1.5\times10^{-33} $\\ \hline
\{3,0\} & 4.481689070 & 4.481689070 & $1.2\times10^{-13} $\\ \hline
\{3,3\} & 20.08553692 & 20.08553692 & $5.4\times10^{-20}$ \\ \hline
\{3,6\} & 90.01713130 & 90.01713130 & $7.3\times10^{-42}$ \\ \hline
\{3,9\} & 403.4287935 & 403.4287935 & $5.4\times10^{-20} $\\ \hline
\{6,0\} & 20.08553692 & 20.08553692 & $1.2\times10^{-13} $\\ \hline
\{6,3\} & 90.01713130 & 90.01713130 & $2.8\times10^{-35 }$\\ \hline
\{6,6\} & 403.4287935 & 403.4287935 & $5.4\times10^{-20}$ \\ \hline
\{6,9\} & 1808.042414 & 1808.042414 & $1.1\times10^{-19} $\\ \hline
\{9,0\} & 90.01713130 & 90.01713130 & $4.6\times10^{-29}$ \\ \hline
\{9,3\} & 403.4287935 & 403.4287935 & $1.2\times10^{-13}$ \\ \hline
\{9,6\} & 1808.042414 & 1808.042414 & $1.2\times10^{-13}$ \\ \hline
\{9,9\} & 8103.083928 & 8103.083928 & $1.2\times10^{-13}$ \\ \hline
\end{tabular}
\vspace{.5cm}
\caption{\label{tab:exp22}Table of values obtained using CA around $(3,6)$.}
\end{subtable}
\end{table}

\subsection{\boldmath $\sin\left(\frac{x+y}{2}\right)$}
We  obtain the $[10/10]$ CA for $\sin\left(\frac{x+y}{2}\right)$ around $(0,0)$ as follows
\begin{mmaCell}{Input}
ChisholmD[Normal[Series[Sin[(x+y)/2],\{x,0,20\},\{y,0,20\}]],\{0,0,10\},\{x,y\}]
\end{mmaCell}
Similarly, We also obtain the $[10/10]$ CA for $\sin\left(\frac{x+y}{2}\right)$ around $(1.6,1.6)$ 
\begin{mmaCell}{Input}
ChisholmD[Normal[Series[Sin[(x+y)/2],\{x,0,20\},\{y,0,20\}]],\{1.6,1.6,10\},\{x,y\}]
\end{mmaCell}
We compare the values obtained using the CA and the values obtained using the in-built \mt function in Table \eqref{tab:sin00} and \eqref{tab:sin1616}. We observe that unlike the case of $\exp\left(\frac{x+y}{2}\right)$ the agreement between the CA and the exact function worsens quickly.

\begin{table}[H]
\caption{Table of values of $\sin\left(\frac{x+y}{2}\right)$}
\begin{subtable}[t]{.5\textwidth}
\raggedright
 \small
\begin{tabular}{|l|l|l|l|}
\hline
\{x,y\}     & CA            & Function      & \% Error                    \\ \hline
\{0.1,0.1\} & 0.09983341665 & 0.09983341665 & $2.4\times10^{-41}$ \\ \hline
\{0.1,1.6\} & 0.7512804051  & 0.7512804051  & $2.7\times10^{-20}$ \\ \hline
\{0.1,3.1\} & 0.9995736030  & 0.9995736030  & $2.0\times10^{-14}$ \\ \hline
\{0.1,4.6\} & 0.7114733528  & 0.7114733528  & $1.0\times10^{-10}$ \\ \hline
\{1.6,0.1\} & 0.7512804051  & 0.7512804051  & $2.7\times10^{-20}$ \\ \hline
\{1.6,1.6\} & 0.9995736030  & 0.9995736030  & $1.2\times10^{-14}$ \\ \hline
\{1.6,3.1\} & 0.7114733528  & 0.7114733528  & $7.3\times10^{-11}$ \\ \hline
\{1.6,4.6\} & 0.04158066227 & 0.04158066243 & $4.0\times10^{-7}$  \\ \hline
\{3.1,0.1\} & 0.9995736030  & 0.9995736030  & $2.0\times10^{-14}$ \\ \hline
\{3.1,1.6\} & 0.7114733528  & 0.7114733528  & $7.3\times10^{-11}$ \\ \hline
\{3.1,3.1\} & 0.04158066201 & 0.04158066243 & $1.0\times10^{-6}$  \\ \hline
\{3.1,4.6\} & -0.6506251887 & -0.6506251371 & $7.9\times10^{-6}$  \\ \hline
\{4.6,0.1\} & 0.7114733528  & 0.7114733528  & $1.0\times10^{-10}$ \\ \hline
\{4.6,1.6\} & 0.04158066227 & 0.04158066243 & $4.0\times10^{-7}$  \\ \hline
\{4.6,3.1\} & -0.6506251887 & -0.6506251371 & $7.9\times10^{-6}$  \\ \hline
\{4.6,4.6\} & -0.9936946941 & -0.9936910036 & 0.00037                     \\ \hline
\end{tabular}
\caption{\label{tab:sin00}Table of values obtained using CA around $(0,0)$.}
\end{subtable}
\begin{subtable}[t]{.56\textwidth}
\raggedleft
\small
\begin{tabular}{|l|l|l|l|}
\hline
\{x,y\}     & CA            & Function      & \% Error                    \\ \hline
\{0.1,0.1\} & 0.09983341665 & 0.09983341665 & $4.1\times10^{-11}$ \\ \hline
\{0.1,1.6\} & 0.7512804051  & 0.7512804051  & $3.8\times10^{-20}$ \\ \hline
\{0.1,3.1\} & 0.9995736030  & 0.9995736030  & $1.9\times10^{-14}$ \\ \hline
\{0.1,4.6\} & 0.7114733528  & 0.7114733528  & $1.1\times10^{-10}$ \\ \hline
\{1.6,0.1\} & 0.7512804051  & 0.7512804051  & $3.8\times10^{-20}$ \\ \hline
\{1.6,1.6\} & 0.9995736030  & 0.9995736030  & $2.4\times10^{-20}$ \\ \hline
\{1.6,3.1\} & 0.7114733528  & 0.7114733528  & $1.8\times10^{-14}$ \\ \hline
\{1.6,4.6\} & 0.04158066243 & 0.04158066243 & $1.1\times10^{-9}$  \\ \hline
\{3.1,0.1\} & 0.9995736030  & 0.9995736030  & $1.9\times10^{-14}$ \\ \hline
\{3.1,1.6\} & 0.7114733528  & 0.7114733528  & $1.8\times10^{-14}$ \\ \hline
\{3.1,3.1\} & 0.04158066243 & 0.04158066243 & $1.0\times10^{-10}$ \\ \hline
\{3.1,4.6\} & -0.6506251369 & -0.6506251371 & $2.3\times10^{-8}$  \\ \hline
\{4.6,0.1\} & 0.7114733528  & 0.7114733528  & $1.1\times10^{-10}$ \\ \hline
\{4.6,1.6\} & 0.04158066243 & 0.04158066243 & $1.1\times10^{-9}$  \\ \hline
\{4.6,3.1\} & -0.6506251369 & -0.6506251371 & $2.3\times10^{-8}$  \\ \hline
\{4.6,4.6\} & -0.9936908505 & -0.9936910036 & 0.000015                    \\ \hline
\end{tabular}
\caption{\label{tab:sin1616} Table of values obtained using CA around $(1.6,1.6)$.}
\end{subtable}
\end{table}

\subsection{\boldmath $\sinh(\frac{x+y}{2})$}
Next, we consider the hyperbolic function 
$\sinh\left(\frac{x+y}{2}\right)$ and find its $[10/10]$ CA around $(0,0)$
\begin{mmaCell}{Input}
ChisholmD[Normal[Series[Sinh[(x+y)/2],\{x,0,20\},\{y,0,20\}]],\{0,0,10\},\{x,y\}]
\end{mmaCell}
Analogously, the $[10/10]$ CA around $(1.6,1.6)$ is obtained as
\begin{mmaCell}{Input}
ChisholmD[Normal[Series[Sinh\Big[\mmaFrac{x+y}{2}\Big],\{x,0,20\},\{y,0,20\}]],\{1.6,1.6,10\},\{x,y\}]
\end{mmaCell}

We compare the values obtained using the CA and the values obtained using the in-built \mt function in Table \eqref{tab:sinh00} and \eqref{tab:sinh1616}. The behaviour of CA for $\sinh(\frac{x+y}{2})$ is similar to that of $\sin(\frac{x+y}{2})$

\begin{table}[h]
\caption{Table of values of $\sinh\left(\frac{x+y}{2}\right)$.}
\begin{subtable}[t]{.5\textwidth}
\raggedright
\small
\begin{tabular}{|l|l|l|l|}
\hline
\{x,y\}     & CA           & Function     & \% Error                    \\ \hline
\{0.1,0.1\} & 0.1001667500 & 0.1001667500 & $2.4\times10^{-41}$ \\ \hline
\{0.1,1.6\} & 0.9561159600 & 0.9561159600 & $2.2\times10^{-20}$ \\ \hline
\{0.1,3.1\} & 2.375567953  & 2.375567953  & $1.0\times10^{-14}$ \\ \hline
\{0.1,4.6\} & 5.195100281  & 5.195100281  & $2.1\times10^{-11}$ \\ \hline
\{1.6,0.1\} & 0.9561159600 & 0.9561159600 & $2.2\times10^{-20}$ \\ \hline
\{1.6,1.6\} & 2.375567953  & 2.375567953  & $5.3\times10^{-15}$ \\ \hline
\{1.6,3.1\} & 5.195100281  & 5.195100281  & $1.2\times10^{-11}$ \\ \hline
\{1.6,4.6\} & 11.07645104  & 11.07645104  & $2.3\times10^{-9}$  \\ \hline
\{3.1,0.1\} & 2.375567953  & 2.375567953  & $1.0\times10^{-14}$ \\ \hline
\{3.1,1.6\} & 5.195100281  & 5.195100281  & $1.2\times10^{-11}$ \\ \hline
\{3.1,3.1\} & 11.07645104  & 11.07645104  & $5.3\times10^{-9}$  \\ \hline
\{3.1,4.6\} & 23.48589183  & 23.48589175  & $3.6\times10^{-7}$  \\ \hline
\{4.6,0.1\} & 5.195100281  & 5.195100281  & $2.1\times10^{-11 }$\\ \hline
\{4.6,1.6\} & 11.07645104  & 11.07645104  & $2.3\times10^{-9}$  \\ \hline
\{4.6,3.1\} & 23.48589183  & 23.48589175  & $3.6\times10^{-7}$  \\ \hline
\{4.6,4.6\} & 49.73713860  & 49.73713190  & 0.000013                    \\ \hline
\end{tabular}
\caption{\label{tab:sinh00} Table of values obtained using CA around the point $(0,0)$.}
\end{subtable}
\begin{subtable}[t]{.55\textwidth}
\raggedleft
\small
\begin{tabular}{|l|l|l|l|}
\hline
\{x,y\}     & CA           & Function     & \% Error                    \\ \hline
\{0.1,0.1\} & 0.1001667500 & 0.1001667500 & $6.1\times10^{-13}$ \\ \hline
\{0.1,1.6\} & 0.9561159600 & 0.9561159600 & $5.6\times10^{-21}$ \\ \hline
\{0.1,3.1\} & 2.375567953  & 2.375567953  & $8.3\times10^{-15 }$\\ \hline
\{0.1,4.6\} & 5.195100281  & 5.195100281  & $1.5\times10^{-11}$ \\ \hline
\{1.6,0.1\} & 0.9561159600 & 0.9561159600 & $5.6\times10^{-21}$ \\ \hline
\{1.6,1.6\} & 2.375567953  & 2.375567953  & $2.1\times10^{-20}$ \\ \hline
\{1.6,3.1\} & 5.195100281  & 5.195100281  & $5.3\times10^{-15}$ \\ \hline
\{1.6,4.6\} & 11.07645104  & 11.07645104  & $1.0\times10^{-11}$ \\ \hline
\{3.1,0.1\} & 2.375567953  & 2.375567953  & $8.3\times10^{-15}$ \\ \hline
\{3.1,1.6\} & 5.195100281  & 5.195100281  & $5.3\times10^{-15}$ \\ \hline
\{3.1,3.1\} & 11.07645104  & 11.07645104  & $9.2\times10^{-16}$ \\ \hline
\{3.1,4.6\} & 23.48589175  & 23.48589175  & $1.3\times10^{-11}$ \\ \hline
\{4.6,0.1\} & 5.195100281  & 5.195100281  & $1.5\times10^{-11}$ \\ \hline
\{4.6,1.6\} & 11.07645104  & 11.07645104  & $1.0\times10^{-11}$ \\ \hline
\{4.6,3.1\} & 23.48589175  & 23.48589175  & $1.3\times10^{-11}$ \\ \hline
\{4.6,4.6\} & 49.73713191  & 49.73713190  & $1.0\times10^{-8}$  \\ \hline
\end{tabular}
\caption{\label{tab:sinh1616}Table of values obtained using CA obtained around the point $(1.6,1.6)$.}
\end{subtable}
\end{table}

\subsection{\boldmath $\log(1+x+y)$}
The series of $\log(1+x+y)$ is given by 
\begin{equation}\label{eq:log00}
    \log(1+x+y) = \sum_{m=1}^{\infty} \frac{(-1)^{m+1}(x+y)^{m}}{m}
\end{equation}
which converges for $|x+y|< 1$. To satisfy the assumption that $a_{00}=1$ is we artificially add 1 to the above series of $\log(1+x+y)$ so as to obtain its CA. To obtain the CA for $\log\left(1+x+y\right)$ around $(0,0)$ we use following command
\begin{mmaCell}{Input}
ChisholmD[1+Normal[Series[Log[1+x+y],\{x,0,20\},\{y,0,20\}]],\{0,0,10\},\{x,y\}]-1
\end{mmaCell} 

\begin{table}[H]
\centering
\begin{tabular}{|l|l|l|l|}
\hline
\{x,y\}     & CA           & Function     & \% Error                    \\ \hline
\{0.1,0.1\} & 0.1823215568 & 0.1823215568 & $5.8\times10^{-31}$ \\ \hline
\{0.1,1.1\} & 0.7884573604 & 0.7884573604 & $4.6\times10^{-14}$ \\ \hline
\{0.1,2.1\} & 1.163150810  & 1.163150810  & $2.4\times10^{-10}$ \\ \hline
\{0.1,3.1\} & 1.435084525  & 1.435084525  & $1.8\times10^{-8}$  \\ \hline
\{1.1,0.1\} & 0.7884573604 & 0.7884573604 & $4.6\times10^{-14}$ \\ \hline
\{1.1,1.1\} & 1.163150810  & 1.163150810  & $3.5\times10^{-12}$ \\ \hline
\{1.1,2.1\} & 1.435084525  & 1.435084525  & $2.3\times10^{-10}$ \\ \hline
\{1.1,3.1\} & 1.648658626  & 1.648658626  & $5.2\times10^{-9}$  \\ \hline
\{2.1,0.1\} & 1.163150810  & 1.163150810  & $2.4\times10^{-10}$ \\ \hline
\{2.1,1.1\} & 1.435084525  & 1.435084525  & $2.3\times10^{-10}$ \\ \hline
\{2.1,2.1\} & 1.648658619  & 1.648658626  & $3.8\times10^{-7}$  \\ \hline
\{2.1,3.1\} & 1.824549134  & 1.824549292  & $8.7\times10^{-6 }$ \\ \hline
\{3.1,0.1\} & 1.435084525  & 1.435084525  & $1.8\times10^{-8}$  \\ \hline
\{3.1,1.1\} & 1.648658626  & 1.648658626  & $5.2\times10^{-9}$  \\ \hline
\{3.1,2.1\} & 1.824549134  & 1.824549292  & $8.7\times10^{-6}$ \\ \hline
\{3.1,3.1\} & 1.974099414  & 1.974081026  & 0.00093                     \\ \hline
\end{tabular}
\caption{\label{tab:log00} Table of values of $\log(1+x+y)$.}
\end{table}

In Table \eqref{tab:log00}, apart from the first entry all the other points lie outside the region of convergence of the series given by Eq. \eqref{eq:log00}.
We thus observe from Table \eqref{tab:log00}, that the CA formed using the series \eqref{eq:log00}, is also valid where the series is not and also matches well with the values obtained using the in-built \mt function (which automatically uses the suitable analytic continuations of $\log(1+x+y)$). The matching worsens as the chosen point moves away from the point $(0,0)$ as is evident from the table. Though with increasing the order of the CA we can obtain better agreement. It is also important to note that the CA obtained using Eq.\eqref{eq:log00} cannot be used for evaluating $\log(1+x+y)$ when the point $(x,y)$ lies on the cut. This is due to the fact that such an approximant does not contain any information of the cut structure of $\log(1+x+y)$ and its suitable analytic continuation should be used to construct CA that would give the correct values on the cut too.

\section{Applications}\label{sec:application}
Analogous to the Pad\'e approximants we study the use of Chisholm approximants for the analytic continuation purposes. As an application of numerical analytic continuation, we consider Appell $F_{1}$\cite{appell1926fonctions,olsson1964integration}, Appell $F_{2}$ \cite{appell1926fonctions,olsson1977integration,Ananthanarayan:2021bqz} and $\text{Li}_{2,2}(x,y)$\cite{kummer1840about,poincare1884groups,chen1977iterated,goncharov2001multiple,Goncharov:2010jf}. However, it is to be noted that the order to which the approximation is taken, affects the numerical value. We  show that the values obtained using approximation are in good agreement with the values obtained by the numerical evaluation of known analytic continuations. 

\subsection{Appell \boldmath $F_1$}\label{appellf1}

We consider double variable Appell $F_1$ series. The analytic continuations of $F_1$ have been previously derived in \cite{olsson1964integration}.  Appell $F_{1}$ is defined as follows \cite{Srivastava:1985,olsson1964integration}
\begin{align}\label{f1definition}
    F_{1}(a,b_{1},b_{2},c,x,y)=\sum_{m, n=0}^{\infty} \frac{(a)_{m+n}\left(b_1\right)_m\left(b_2\right)_n}{(c)_{m+n} m ! n !} x^m y^n
\end{align}
with region of convergence : $ |x|< 1 \wedge |y| < 1 $. We discuss various properties of CA obtained for this series below.

We form $[10/10]$ approximant of the above series by taking terms from $m,n= 0 \cdots 20$\footnote{We remark that such a sum is taken for the convenient purposes. For $[10/10]$ approximant only 251 terms are required.}. The comparison of values obtained using $[10/10]$ CA and values obtained by summing the series ( Eq.\eqref{f1definition}) from 0 to 100 in each of the summation indices,  are shown in Table \eqref{tab:f200}. Furthermore, we have the following transformation formula of $F_{1}$
\begin{equation}\label{f1ac}
  F_{2}(a,b_{1},b_{2},c,x,y)=  (1-x)^{b_{1}} (1-y)^{b_2}  F_{2}\left(c-a,b_{1},b_{2},c,\frac{x}{x-1},\frac{y}{y-1}\right)
\end{equation}
where the RHS converges for: $\dfrac{x}{x-1}< 1 \wedge \dfrac{y}{y-1}< 1$. This relation provides the analytic continuation of $F_{1}$ and covers the whole third quadrant of the real $x-y$ plane.

\begin{figure}[H]
    \centering
    \includegraphics[width=6.5cm, height=6.5cm]{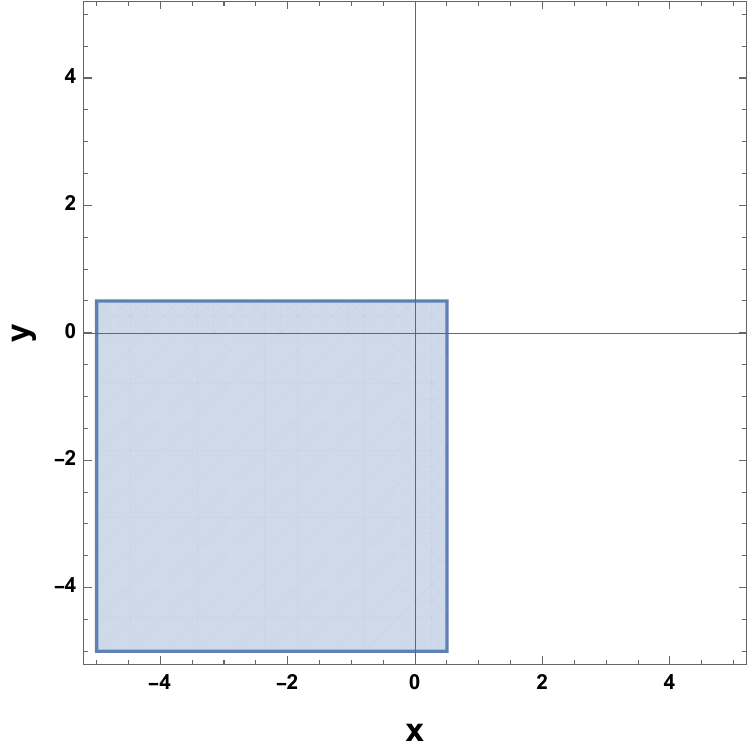}
    \caption{ROC of AC of $F_2$ given by Eq.\eqref{f1ac}.}
    \label{fig:f1ac}
\end{figure}

 In Table\eqref{tab:f2ab} we compare the values obtained using $[10/10]$ CA of Eq.\eqref{f1definition} and values obtained by summing the series in Eq.\eqref{f1ac} from 0 to 100 in each of the summation indices. We see that the values obtained using CA matches even outside the ROC of Eq.\eqref{f1definition}.

\begin{table}[h]
\caption{Comparison of values obtained using CA and the function $F_{1}\left(a,b_{1},b_{2},c,x,y\right)$ , Eq.\eqref{f1definition}.}
\begin{subtable}[t]{.5\textwidth}
\raggedright
\small
\begin{tabular}{|l|l|l|l|}
\hline
\{x,y\}       & CA          & Function    & \% Error                                             \\ \hline
\{0.1,0.1\}   & 1.207502432 & 1.207502432 & $6.1 \times 10^{-24}$                                \\ \hline
\{0.1,0.34\}  & 1.471358983 & 1.471358983 & $2.3\times 10^{-15}$                                 \\ \hline
\{0.1,0.58\}  & 1.948702367 & 1.948702367 & $1.1\times 10^{-10}$                                 \\ \hline
\{0.1,0.82\}  & 3.245962293 & 3.245962139 & $4.7\times 10^{-6}$                                  \\ \hline
\{0.34,0.1\}  & 1.659460805 & 1.659460805 & $1.8\times 10^{-15}$                                 \\ \hline
\{0.34,0.34\} & 1.961119271 & 1.961119271 & $5.3\times 10^{-11}$                                 \\ \hline
\{0.34,0.58\} & 2.502849864 & 2.502849840 & $9.9\times 10^{-7}$                                  \\ \hline
\{0.34,0.82\} & 3.962087028 & 3.961749783 & 0.0085                                               \\ \hline
\{0.58,0.1\}  & 2.530523511 & 2.530523511 & $7.9\times 10^{-11}$ \\ \hline
\{0.58,0.34\} & 2.900515131 & 2.900515140 & $3.4\times 10^{-7}$                                  \\ \hline
\{0.58,0.58\} & 3.557119523 & 3.557105989 & 0.00038                                              \\ \hline
\{0.58,0.82\} & 5.269593568 & 5.298264613 & 0.54                                                 \\ \hline
\{0.82,0.1\}  & 5.155409814 & 5.155410035 & $4.3\times 10^{-6}$                                  \\ \hline
\{0.82,0.34\} & 5.715281448 & 5.715371394 & 0.0016                                               \\ \hline
\{0.82,0.58\} & 6.686208931 & 6.684782078 & 0.021                                                \\ \hline
\end{tabular}
\caption{\label{tab:f200}Table of values obtained using CA of Eq.\eqref{f1definition}.}
\end{subtable}
\begin{subtable}[t]{.55\textwidth}
\raggedleft
\small
\begin{tabular}{|l|l|l|l|}
\hline
\{x,y\}       & CA             & Eq.\eqref{f1ac}       & \% Error \\ \hline
\{-1.,-1.\}   & 0.07863469382  & 0.07863466908  & 0.000031 \\ \hline
\{-1.,-1.5\}  & 0.009242093487 & 0.009242025192 & 0.00074  \\ \hline
\{-1.,-2.\}   & -0.04009680166 & -0.04009609854 & 0.0018   \\ \hline
\{-1.,-2.5\}  & -0.07715771667 & -0.07715280459 & 0.0064   \\ \hline
\{-1.,-3.\}   & -0.1061062366  & -0.1060897986  & 0.015    \\ \hline
\{-1.5,-1.\}  & -0.03466561327 & -0.03466663472 & 0.0029   \\ \hline
\{-1.5,-1.5\} & -0.09715488084 & -0.09716351628 & 0.0089   \\ \hline
\{-1.5,-2.\}  & -0.1413096277  & -0.1413367821  & 0.019    \\ \hline
\{-1.5,-2.5\} & -0.1742764748  & -0.1743217876  & 0.026    \\ \hline
\{-1.5,-3.\}  & -0.1998862081  & -0.1999311654  & 0.022    \\ \hline
\{-2.,-1.\}   & -0.1120107537  & -0.1120225723  & 0.011    \\ \hline
\{-2.,-1.5\}  & -0.1693922192  & -0.1695202070  & 0.076    \\ \hline
\{-2.,-2.\}   & -0.2095106666  & -0.2099700949  & 0.22     \\ \hline
\{-2.,-2.5\}  & -0.2392184278  & -0.2400338776  & 0.34     \\ \hline
\{-2.,-3.\}   & -0.2622159926  & -0.2632658502  & 0.40     \\ \hline
\end{tabular}
\caption{\label{tab:f2ab} Comparison of values obtained using CA of Eq.\eqref{f1definition} and the values obtained using AC, Eq.\eqref{f1ac}.}
\end{subtable}
\end{table}

\subsection{Appell \boldmath $F_2$}
Appell $F_{2}$ is defined as \cite{Srivastava:1985,olsson1977integration,Ananthanarayan:2021bqz}
\begin{equation}\label{appellf2}
   F_{2}(a,b_{1},b_{2},c_{1},c_{2},x,y)= \sum_{m, n=0}^{\infty} \frac{(a)_{m+n}\left(b_1\right)_m\left(b_2\right)_n}{(c_1)_{m}(c_2)_{n} m ! n !} x^m y^n
\end{equation}
which converges for $|x|+|y| <1$.
We compute the $[10/10]$ CA of Appell $F_{2}$ with the pochhammer parameters: $a=\frac{3}{10}, b_1=\frac{4}{10}, b_2=\frac{3}{17}, c_1=\frac{1}{5}, c_2=\frac{1}{7}$. To find the CA we take series of $F_{2}$ from $m,n =0, \cdots, 20$. We tabulate the comparison of values obtained using the CA around $(0,0)$ and using the series, Eq.\ref{appellf2} in table \eqref{tab:f200}. In table \eqref{tab:f2ab} we compare values obtained using the CA around $(x,y)$(the point at which the result is required) with the values obtained using the package \texttt{AppellF2.wl}. The values from the package \ft as well as the values obtained by summing Eq.\eqref{appellf2} are obtained in both cases by summing the series from 0 to 150 in each of the summation indices. We observe from the tables \eqref{tab:f200} and \eqref{tab:f2ab} that the agreement between the values using CA and values obtained using \texttt{AppellF2.wl} is better when the points lie in the first and third quadrant. 
\begin{table}[h]
\caption{Table of values of Appell $F_{2}$}
\begin{subtable}[t]{.5\textwidth}
\raggedright
\small
\begin{tabular}{|l|l|l|l|}
\hline
\{x,y\}       & CA           & Eq.\eqref{appellf2}     & Error                      \\ \hline
\{-0.6,-0.2\} & 0.7373422441 & 0.7364876835 & 0.12                       \\ \hline
\{-0.6,0.2\}  & 0.7596772196 & 0.7594320572 & 0.032                      \\ \hline
\{-0.2,-0.6\} & 0.7889825335 & 0.7891964716 & 0.027                      \\ \hline
\{-0.2,-0.2\} & 0.8549285608 & 0.8549285555 & $6.2\times 10^{-7}$ \\ \hline
\{-0.2,0.2\}  & 0.9431865106 & 0.9431860672 & 0.000047                   \\ \hline
\{-0.2,0.6\}  & 1.063513188  & 1.059374908  & 0.39                       \\ \hline
\{0.2,-0.6\}  & 0.8959261238 & 0.8960573837 & 0.015                      \\ \hline
\{0.2,-0.2\}  & 1.035523120  & 1.035523421  & 0.000029                   \\ \hline
\{0.2,0.2\}   & 1.298900246  & 1.298899102  & 0.000088                   \\ \hline
\{0.6,-0.2\}  & 1.355309400  & 1.356644887  & 0.098                      \\ \hline
\{0.6,0.2\}   & -2.473368787 & 2.533662025  & $2.0\times 10^{2}$  \\ \hline
\end{tabular}
\caption{\label{tab:f200} Table of values obtained using CA around $(0,0)$ of Eq.\eqref{appellf2}.}
\end{subtable}
\begin{subtable}[t]{.55\textwidth}
\raggedleft
\small
\begin{tabular}{|l|l|l|l|}
\hline
\{x,y\}       & CA           & Eq.\eqref{eq:f2ac}     & Error                       \\ \hline
\{-0.6,-0.2\} & 0.7364873706 & 0.7364876835 & 0.000042                    \\ \hline
\{-0.6,0.2\}  & 0.7597201054 & 0.7594320572 & 0.038                       \\ \hline
\{-0.2,-0.6\} & 0.7891961013 & 0.7891964716 & 0.000047                    \\ \hline
\{-0.2,-0.2\} & 0.8549285555 & 0.8549285555 & $1.0\times 10^{-14}$ \\ \hline
\{-0.2,0.2\}  & 0.9431860672 & 0.9431860672 & $4.1\times 10^{-12}$ \\ \hline
\{-0.2,0.6\}  & 1.059376291  & 1.059374908  & 0.00013                     \\ \hline
\{0.2,-0.6\}  & 0.8962813844 & 0.8960573837 & 0.025                       \\ \hline
\{0.2,-0.2\}  & 1.035523421  & 1.035523421  & $2.9\times 10^{-12}$ \\ \hline
\{0.2,0.2\}   & 1.298899102  & 1.298899102  & $8.7\times 10^{-12}$ \\ \hline
\{0.6,-0.2\}  & 1.356646128  & 1.356644887  & 0.000091                    \\ \hline
\{0.6,0.2\}   & 2.531769261  & 2.533662025  & 0.075                       \\ \hline
\end{tabular}
\caption{\label{tab:f2ab}Table of values obtained CA around $(x,y)$ of Eq.\eqref{appellf2}.}
\end{subtable}
\end{table}

Outside its region of convergence, $F_2$ can be evaluated using the analytic continuations(ACs) evaluated in \cite{Ananthanarayan:2021bqz}. To illustrate the use of the package and to obtain the CA when the series is not of the form given by Eq.\eqref{eq:genseries}, we take the following AC of Appell $F_2$ \cite{Ananthanarayan:2021bqz} 
\begin{align}\label{eq:f2ac}
& F_{2}(a,b_{1},b_{2},c_{1},c_{2},x,y) = \frac{\Gamma \left(c_2\right)\Gamma \left(b_2-a\right)}{\Gamma \left(b_2\right) \Gamma \left(c_2-a\right)} (-y)^{-a} \sum_{m,n=0}^{\infty}\frac{\left(b_1\right)_m (a)_{m+n} \left(a-c_2+1\right)_{m+n}}{m! n!  \left(c_1\right)_m \left(a-b_2+1\right)_{m+n}} \left(-\frac{x}{y}\right)^{m}\left(\frac{1}{y}\right)^{n} + \nonumber \\
 &\frac{\Gamma \left(c_1\right) \Gamma \left(c_2\right) \Gamma \left(a-b_1-b_2\right)(-x)^{-b_1} (-y)^{-b_2}}{\Gamma (a) \Gamma \left(c_1-b_1\right) \Gamma \left(c_2-b_2\right)} \sum_{m,n=0}^{\infty} \frac{\left(b_1\right)_m \left(b_2\right)_n \left(b_1-c_1+1\right)_m \left(b_2-c_2+1\right)_n}{\left(-a+b_1+b_2+1\right)_{m+n} m! n!} \left(\frac{1}{x}\right)^{m}\left(\frac{1}{y}\right)^{n}+ (-x)^{b_2-a}\nonumber \\
 & \frac{ \Gamma (c_1) \Gamma(c_2)\Gamma(a-b_2) \Gamma(-a+b_1+b_2)(-y)^{-b_2}}{\Gamma (a) \Gamma(b_1) \Gamma(c_2-b_2) \Gamma(-a+b_2+c_1)} \sum_{m,n=0}^{\infty}\frac{(b_2)_n (b_2-c_2+1)_n (a-b_2)_{m-n} (a-b_2-c_1+1)_{m-n}}{(a-b_1-b_2+1)_{m-n} m! n!} \left(\frac{1}{x}\right)^{m}\left(\frac{x}{y}\right)^{n}
\end{align}
which converges for : $\frac{1}{| x| }<1\land \left| \frac{x}{y}\right| <1\land \left| \frac{x}{y}\right| < \left|\frac{ x }{ x +1}\right|\land \frac{1}{| y| }<1\land \left| \frac{x}{y}\right| +\frac{1}{| y| }<1$.

It is to be noted that the series in Eq.\eqref{eq:f2ac} is not of the form given by Eq.\eqref{eq:genseries}. To find the CA for these it is first desirable to convert the series into the form given by Eq.\eqref{eq:genseries}. We simply do this by considering the series of the following form 
\begin{equation*}
    \sum_{m,n=0}^{\infty} c_{mn}X^{m}Y^{n}
\end{equation*}
Here $X$ and $Y$ are function of $x$ and $y$. As an example the first series in Eq.\eqref{eq:f2ac} has:
\begin{equation*}
    c_{mn}= \frac{\left(b_1\right)_m (a)_{m+n} \left(a-c_2+1\right)_{m+n}}{m! n!  \left(c_1\right)_m \left(a-b_2+1\right)_{m+n}} \quad X= -\frac{x}{y} \quad Y= \frac{1}{y}
\end{equation*}
We then obtain its CA using the following command
\begin{mmaCell}{Input}
ChisholmD[series1,\{0,0,10\},\{X,Y\}]
\end{mmaCell}
here, \texttt{series1} denotes the first series in Eq.\eqref{eq:f2ac}. In the CA thus obtained, as a function of $X$ and $Y$, we then substitute $X= -\frac{x}{y}$ and $Y= \frac{1}{y}$ so as to obtain the CA for the first series in Eq.\eqref{eq:f2ac}. Similarly, the procedure can be repeated for other series in Eq.\eqref{eq:f2ac}.

In Table \eqref{table:f2ac} we present the values obtained using the CA of Eq.\eqref{eq:f2ac} and \ft. The first column denotes whether the point at which Appell $F_{2}$ is evaluated lies inside the ROC of Eq.\eqref{eq:f2ac} or not by means of True and False respectively.

\begin{table}[t]
\centering
\begin{tabular}{|l|l|l|l|l|}
\hline
ROC   & \{x,y\}       & CA                                                                        & \ft o/p                                                                & \% Error                      \\ \hline
False & \{5.,5.\}     & \begin{tabular}[c]{@{}l@{}}-0.10488036372\\ +0.08296505721 I\end{tabular} & \begin{tabular}[c]{@{}l@{}}-0.10487887277\\ +0.08296374276 I\end{tabular} & 0.0015                     \\ \hline
True  & \{5.,15.\}    & \begin{tabular}[c]{@{}l@{}}-0.04109494944\\ +0.03474758530 I\end{tabular} & \begin{tabular}[c]{@{}l@{}}-0.04109494941\\ +0.03474758527 I\end{tabular} & $8.5\times 10^{-8}$ \\ \hline
False & \{15.,5.\}    & \begin{tabular}[c]{@{}l@{}}-0.04051289211\\ +0.03402313454 I\end{tabular} & \begin{tabular}[c]{@{}l@{}}-0.04050813057\\ +0.03401893668 I\end{tabular} & 0.012                      \\ \hline
False & \{15.,15.\}   & \begin{tabular}[c]{@{}l@{}}-0.02173711582\\ +0.01889956754 I\end{tabular} & \begin{tabular}[c]{@{}l@{}}-0.02173711133\\ +0.01889956359 I\end{tabular} & 0.000021                   \\ \hline
False & \{-15.,5.\}   & \begin{tabular}[c]{@{}l@{}}-0.6992795762\\ -0.2380483933 I\end{tabular}   & \begin{tabular}[c]{@{}l@{}}0.07814958058\\ -0.05575437776 I\end{tabular}  & $8.3\times 10^{2}$  \\ \hline
False & \{-15.,15.\}  & \begin{tabular}[c]{@{}l@{}}0.2437754672\\ -0.5199484573 I\end{tabular}    & \begin{tabular}[c]{@{}l@{}}0.0432580097\\ -0.1778996098 I\end{tabular}    & $2.2\times 10^{2}$  \\ \hline
False & \{-5.,5.\}    & \begin{tabular}[c]{@{}l@{}}-0.3338565813\\ -0.6223061875 I\end{tabular}   & \begin{tabular}[c]{@{}l@{}}0.2212243174\\ -0.4558571901 I\end{tabular}    & $1.1\times 10^{2}$  \\ \hline
True  & \{-5.,15.\}   & \begin{tabular}[c]{@{}l@{}}-0.11559676346\\ -0.01973305678 I\end{tabular} & \begin{tabular}[c]{@{}l@{}}-0.11559676363\\ -0.01973305706 I\end{tabular} & $2.8\times 10^{-7}$ \\ \hline
False & \{-15.,-15.\} & 0.02446537613                                                             & 0.02446537612                                                             & $3.4\times 10^{-8}$ \\ \hline
False & \{-15.,-5.\}  & 0.04005221085                                                             & 0.04005451644                                                             & 0.0058                     \\ \hline
True  & \{-5.,-15.\}  & 0.04088637870                                                             & 0.04088637870                                                             & $5.4\times 10^{-9}$ \\ \hline
False & \{-5.,-5.\}   & 0.08293654594                                                             & 0.08293657495                                                             & 0.000035                   \\ \hline
True  & \{5.,-15.\}   & \begin{tabular}[c]{@{}l@{}}0.08472607959\\ -0.06470315121 I\end{tabular}  & \begin{tabular}[c]{@{}l@{}}0.08472607937\\ -0.06470315236 I\end{tabular}  & $1.1\times 10^{-6}$ \\ \hline
False & \{5.,-5.\}    & \begin{tabular}[c]{@{}l@{}}0.1830816882\\ -0.3574591665 I\end{tabular}    & \begin{tabular}[c]{@{}l@{}}0.1480028697\\ -0.5418921340 I\end{tabular}    & 33.                        \\ \hline
False & \{15.,-15.\}  & \begin{tabular}[c]{@{}l@{}}-0.0018909199\\ -0.1974597530 I\end{tabular}   & \begin{tabular}[c]{@{}l@{}}-0.0018989046\\ -0.1974707931 I\end{tabular}   & 0.0069                     \\ \hline
False & \{15.,-5.\}   & \begin{tabular}[c]{@{}l@{}}0.3476194581\\ -0.7270318659 I\end{tabular}    & \begin{tabular}[c]{@{}l@{}}-0.10344529632\\ -0.01090328199 I\end{tabular} & $8.1\times 10^{2}$  \\ \hline
\end{tabular}
\caption{Table for values using CA of Eq.\eqref{appellf2} and using the \ft}\label{table:f2ac}
\end{table}

We observe from the above that even when the point is outside the ROC of Eq.\eqref{eq:f2ac} the values obtained using the CA are in good agreement with the value obtained using the package \ft. More specifically this happens when the points lie either in the first quadrant or the third quadrant and we have a mismatch for the values lying in second or fourth quadrant.

\subsection{Application in condensed matter- 1}\label{sec:condeg1}
In \cite{wood1974chisholm,wood1975applications} it is required to obtain the two variable approximants of the zero-field susceptibility of the three-dimensional Ising model. The double series expansion of the susceptibility can be written as follows 
\begin{equation}\label{eqn:condeg1}
    f(z_{1},z_{2})= 1+  \sum_{l} P_{l}(z_{1})z_{2}^{l}
\end{equation}
where $z_{1}$ and $z_{2}$ are functions of coupling constants and temperature, which we omit for the present purpose. $P_{l}(z_{1})$ is a polynomial of degree $l$ in $z_{1}$. In \cite{dalton1969critical} the polynomial $P_{l}(z_{1})$ has been evaluated for various systems. For the illustration purposes of the package, we will take $P_{l}(z_{1})$ to be Legendre polynomials. Though, it is not related to any physical system still it would illustrate the procedure well without loss of its features. It would also be advantageous as we would be able to calculate higher order CA in contrast to \cite{dalton1969critical} where such polynomials are not known up to very high order. We present the result in the accompanying \mt file \texttt{Condensed\_matter.nb} in the section \texttt{Example 1}.

We now describe some of the features of this study. Firstly, we note that for the series given by Eq. \eqref{eqn:condeg1} the CA does not exist as the correct number of consistency equations cannot be formed for a given order. To remove this problem we would instead consider the following transformation of the variables
\begin{equation*}
    z_{1} \to x-y \quad z_{2} \to x+y.
\end{equation*}
After the above transformation, we further need to add $x+y$ to the resulting series to as to obtain CA. Finally from the CA thus obtained we would subtract $x+y$. It can be done using the following command
\begin{mmaCell}{Input}
 ChisholmD[x+y+f(x,y),\{0,0,10\},\{x,y\}]-(x+y)
\end{mmaCell}
where $f(x,y)$ is the Eq.\eqref{eqn:condeg1} after transformation. In the above obtain CA then do the reverse transformation: 
\begin{equation*}
    x \to \frac{z_{1}+z_{2}}{2}, \quad y \to \frac{z_{2}- z_{1}}{2}
\end{equation*}
to obtain the approximants in terms of $z_{1}$ and $z_{2}$.

In Table \eqref{table:condeg1} we compare the values obtained using the CA thus obtained and summing the series given by Eq.\eqref{eqn:condeg1} from $l=0 \cdots 100$.
\begin{table}[H]
\centering
\begin{tabular}{|l|l|l|l|}
\hline
\{x,y\}       & CA           & Eq.\eqref{eqn:condeg1}     & \% Error                       \\ \hline
\{0.01,0.01\} & 1.000050004  & 1.000050004  & $7.7\times 01^{-45}$ \\ \hline
\{0.01,0.21\} & 0.9806278224 & 0.9806278224 & $1.7\times 01^{-17}$ \\ \hline
\{0.01,0.41\} & 0.9285167140 & 0.9285167140 & $1.0\times 01^{-11}$ \\ \hline
\{0.01,0.61\} & 0.8575244528 & 0.8575244529 & $1.6\times 01^{-8}$  \\ \hline
\{0.01,0.81\} & 0.7808926017 & 0.7808926175 & $2.0\times 01^{-6}$  \\ \hline
\{0.21,0.01\} & 1.002056325  & 1.002056325  & $6.9\times 01^{-21}$ \\ \hline
\{0.21,0.21\} & 1.022807183  & 1.022807183  & $7.4\times 01^{-18}$ \\ \hline
\{0.21,0.41\} & 1.002056325  & 1.002056325  & $5.2\times 01^{-15}$ \\ \hline
\{0.21,0.61\} & 0.9466454705 & 0.9466454705 & $7.2\times 01^{-12}$ \\ \hline
\{0.21,0.81\} & 0.8717431760 & 0.8717431762 & $3.1\times 01^{-8}$  \\ \hline
\{0.41,0.01\} & 1.004074771  & 1.004074771  & $6.9\times 01^{-16}$ \\ \hline
\{0.41,0.21\} & 1.070943751  & 1.070943751  & $2.5\times 01^{-12}$ \\ \hline
\{0.41,0.41\} & 1.096388415  & 1.096388415  & $9.7\times 01^{-12}$ \\ \hline
\{0.41,0.61\} & 1.070943751  & 1.070943751  & $1.3\times 01^{-11}$ \\ \hline
\{0.41,0.81\} & 1.004074771  & 1.004074771  & $2.2\times 01^{-9}$  \\ \hline
\{0.61,0.01\} & 1.006105463  & 1.006105463  & $3.5\times 01^{-13}$ \\ \hline
\{0.61,0.21\} & 1.126586259  & 1.126586259  & $2.0\times 01^{-9 }$ \\ \hline
\{0.61,0.41\} & 1.223613551  & 1.223613552  & $5.2\times 01^{-8 }$ \\ \hline
\{0.61,0.61\} & 1.261986642  & 1.261986643  & $9.9\times 01^{-8 }$ \\ \hline
\{0.61,0.81\} & 1.223613551  & 1.223613552  & $9.9\times 01^{-8 }$ \\ \hline
\{0.81,0.01\} & 1.008148527  & 1.008148527  & $2.5\times 01^{-11}$ \\ \hline
\{0.81,0.21\} & 1.191912890  & 1.191912892  & $2.1\times 01^{-7}$  \\ \hline
\{0.81,0.41\} & 1.408729931  & 1.408730186  & 0.000018                    \\ \hline
\{0.81,0.61\} & 1.613950569  & 1.613953225  & 0.00016                     \\ \hline
\{0.81,0.81\} & 1.705228240  & 1.705233720  & 0.00032                     \\ \hline
\end{tabular}
\caption{Table of values obtained using CA of Eq.\eqref{eqn:condeg1} and Eq.\eqref{eqn:condeg1} }\label{table:condeg1}
\end{table}

\subsection{Application in condensed matter- 2}\label{sec:condeg2}
Another example where such approximants have been useful has been in the study of critical phenomena in \cite{watson1974two,wood1974chisholm}. The function that is of interest in these studies is the following
\begin{equation}\label{eqn:condeg2}
    f(z_{1},z_{2}) = \frac{1}{e^{z_{1} z_{2}}-z_{2}}
\end{equation}
We again observe that similar to the example in sub-section \eqref{sec:condeg1}, CA cannot be obtained using the series of the right-hand side of Eq.\eqref{eqn:condeg2}. Repeating the same procedure as described in sub-section \eqref{sec:condeg1} we obtain the CA for Eq.\eqref{eqn:condeg2}. The comparison of values thus obtained are presented in Table \eqref{table:condeg2}. The value of the function is obtained using the in-built \mt function.

\begin{table}[H]
\centering
\begin{tabular}{|l|l|l|l|}
\hline
\{x,y\}       & CA            & Eq.\eqref{eqn:condeg2}      & \% Error                       \\ \hline
\{0.1,0.1\}   & 1.098840521   & 1.098840521   & $1.3\times 01^{-28}$ \\ \hline
\{0.1,1.1\}   & 61.43234252   & 61.43234252   & $3.0\times 01^{-10}$ \\ \hline
\{0.1,2.1\}   & -1.154305335  & -1.154305292  & $3.7\times 01^{-6}$  \\ \hline
\{0.4,0.1\}   & 1.062912997   & 1.062912997   & $2.9\times 01^{-20}$ \\ \hline
\{0.4,1.1\}   & 2.208933189   & 2.208933189   & $1.7\times 01^{-9 }$ \\ \hline
\{0.4,2.1\}   & 4.621756970   & 4.621777384   & 0.00044                     \\ \hline
\{0.7,0.1\}   & 1.028268985   & 1.028268985   & $5.8\times 01^{-16}$ \\ \hline
\{0.7,1.1\}   & 0.9436043051  & 0.9436043056  & $4.9\times 01^{-8}$  \\ \hline
\{0.7,2.1\}   & 0.4445907354  & 0.4445955791  & 0.0011                      \\ \hline
\{1.,0.1\}    & 0.9948556828  & 0.9948556828  & $4.7\times 01^{-13}$ \\ \hline
\{1.,1.1\}    & 0.5251642849  & 0.5251642910  & $1.2\times 01^{-6}$  \\ \hline
\{1.,2.1\}    & 0.1648319239  & 0.1648486631  & 0.010                       \\ \hline
\{1.3,0.1\}   & 0.9626229087  & 0.9626229087  & $7.7\times 01^{-11}$ \\ \hline
\{1.3,1.1\}   & 0.3248124384  & 0.3248125061  & 0.000021                    \\ \hline
\{1.3,2.1\}   & 0.07541934296 & 0.07556929931 & 0.20                        \\ \hline
\{1.6,0.1\}   & 0.9315229375  & 0.9315229375  & $4.7\times 01^{-9}$  \\ \hline
\{1.6,1.1\}   & 0.2122038089  & 0.2122044106  & 0.00028                     \\ \hline
\{1.6,2.1\}   & 0.03776965663 & 0.03746835206 & 0.80                        \\ \hline
\{1.9,0.1\}   & 0.9015103550  & 0.9015103563  & $1.5\times 01^{-7}$  \\ \hline
\{1.9,1.1\}   & 0.1431607395  & 0.1431656615  & 0.0034                      \\ \hline
\{1.9,2.1\}   & 0.01957320339 & 0.01924746664 & 1.7                         \\ \hline
\{0.81,0.21\} & 1.191912890   & 1.191912892   & $2.1\times 01^{-7}$  \\ \hline
\{0.81,0.41\} & 1.408729931   & 1.408730186   & 0.000018                    \\ \hline
\{0.81,0.61\} & 1.613950569   & 1.613953225   & 0.00016                     \\ \hline
\{0.81,0.81\} & 1.705228240   & 1.705233720   & 0.00032                     \\ \hline
\end{tabular}
\caption{Table of values obtained using CA of Eq.\eqref{eqn:condeg2}
 and Eq.\eqref{eqn:condeg2}}\label{table:condeg2}
\end{table}

\subsection{\boldmath $\mathrm{Li}_{2,2}(x, y)$}

The classical polylogarithms are defined by iterative integrals
\begin{align}
    \text{Li}_{n+1} (x) = \int_0^x \frac{\text{Li}_n(t)}{t} dt
\end{align}
It can also be represented as infinite sums
\begin{align}
    \text{Li}_n (x) = \sum_{i=1}^\infty \frac{x^i}{i^n}
\end{align}
which is valid for $|x|<1$. The value of the polylogarithms can be found for $|x|\geq 1$ by performing analytic continuations.

Similarly, $\text{Li}_{2,2}(x,y)$  is defined as 
\begin{align}
    \text{Li}_{2,2}(x,y) = \sum_{i>j>0}^\infty \frac{x^i y^j}{i^2 j^2} = \sum_{i=1,j=1}^\infty \frac{x ^i (x y)^j}{(i+j)^2 j^2} \label{eqn:Li_def}
\end{align}
which is valid in the domain $|x|\leq1 \wedge |x y| \leq 1$. Note that, this order of argument in the above definition is same as that of \cite{Frellesvig:2016ske, Vollinga:2004sn},  but revered compared to the definitions in \cite{goncharov2001multiple,goncharov2011multiple}.
The classical and multiple polylogarithms frequently appear in the Feynman integral calculus. There exist computer programs, that can handle the manipulation and evaluations of MPLs \cite{Duhr:2019tlz,Frellesvig:2018lmm,Vollinga:2004sn, Wang:2021imw,Naterop:2019xaf, Frellesvig:2016ske}.

In \cite{Duhr:2011zq}, it is conjectured that all the MPLs up to weight 4 can be expressed in terms of classical polylogarithms with weight up to 4 and  $\text{Li}_{2,2}(x,y)$, which is later proved in \cite{Frellesvig:2016ske}. Furthermore, the authors of the later paper provides an algorithm to evaluate the double variable series  $\text{Li}_{2,2}(x,y)$. In the remainder of the section we study the CA of of the series of $\text{Li}_{2,2}(x,y)$.

It is worth pointing out that, the bivariate MPLs can be written in terms of Kamp\'e de F\'eriet functions as follows
\begin{align}
    \text{Li}_{p,q}(x,y) = \frac{x^2 y}{2^p}\text{KdF}^{p:1;q+1}_{p:0;q}
  \left[
   \setlength{\arraycolsep}{0pt}
   \begin{array}{c@{{}:{}}c@{;{}}c} 2,\dots,2& 1 & 1,\dots, 1
 \\[1ex]
  3, \dots, 3& \linefill & 2, \dots ,2  
   \end{array}
   \;\middle|\;
 x,x y
 \right] 
\end{align}

The following relations are well known from the literature  which can be used to find the numerical value of $\text{Li}_{2,2}(x,y)$ beyond its defining region of convergence (ROC),
\begin{align}
    \mathrm{Li}_{2,2}(x, y) &=-\mathrm{Li}_{2,2}(y, x)-\mathrm{Li}_4(x y)+\mathrm{Li}_2(x) \mathrm{Li}_2(y) \label{eqn:stuffle}\\
\mathrm{Li}_{2,2}(x, y) & =\mathrm{Li}_{2,2}\left(\frac{1}{x}, \frac{1}{y}\right)-\mathrm{Li}_4(x y)+3\left(\mathrm{Li}_4\left(\frac{1}{x}\right)+\mathrm{Li}_4(y)\right)+2\left(\mathrm{Li}_3\left(\frac{1}{x}\right)-\mathrm{Li}_3(y)\right) \log (-x y) \nonumber\\
& +\mathrm{Li}_2\left(\frac{1}{x}\right)\left(\frac{\pi^2}{6}+\frac{\log ^2(-x y)}{2}\right)+\frac{1}{2} \mathrm{Li}_2(y)\left(\log ^2(-x y)-\log ^2(-x)\right)   \label{eqn:inversion}
\end{align}
The first of the two relations (i.e., Eq. \eqref{eqn:stuffle}) is known as the stuffle relation, and the second relation is known as the inversion relation. 
Another relation can be obtained by applying the stuffle relation on the $\mathrm{Li}_{2,2}\left(\frac{1}{x}, \frac{1}{y}\right)$ appearing on the RHS of Eq. \eqref{eqn:inversion},
\begin{small}

\begin{align}
   \mathrm{Li}_{2,2}(x, y) & =-\mathrm{Li}_{2,2}\left(\frac{1}{y}, \frac{1}{x}\right)-\mathrm{Li}_4\left(\frac{1}{x y}\right)+\mathrm{Li}_2\left(\frac{1}{x})\right) \mathrm{Li}_2\left(\frac{1}{y}\right)-\mathrm{Li}_4(x y)+3\left(\mathrm{Li}_4\left(\frac{1}{x}\right)+\mathrm{Li}_4(y)\right)\\
   &+2\left(\mathrm{Li}_3\left(\frac{1}{x}\right)-\mathrm{Li}_3(y)\right) \log (-x y)  +\mathrm{Li}_2\left(\frac{1}{x}\right)\left(\frac{\pi^2}{6}+\frac{\log ^2(-x y)}{2}\right)+\frac{1}{2} \mathrm{Li}_2(y)\left(\log ^2(-x y)-\log ^2(-x)\right)    \label{eqn:stuffleinversion}
\end{align}
    
\end{small}

These relations are valid for
\begin{equation}
\label{eq:rocLi}
\begin{aligned}
    \text{Eq.} \eqref{eqn:Li_def} &:  \hspace{1cm}|x|<1 \wedge |x y| <1\\
    \text{Eq.} \eqref{eqn:stuffle} &: \hspace{1cm}|y| < 1 \wedge |x y| < 1 \\
    \text{Eq.} \eqref{eqn:inversion} &: \hspace{1cm} \left| x \right| > 1\land \left| x y \right| > 1\\
    \text{Eq.} \eqref{eqn:stuffleinversion} &: \hspace{1cm} \left| x \right| < 1\land \left| x y \right| > 1
\end{aligned}
\end{equation}

Note that the first few terms of the $\mathrm{Li}_{2,2}(x, y)$  series are
\begin{align*}
    \mathrm{Li}_{2,2}(x, y) = \frac{x^2 y}{4} + \frac{x^3 y}{9} + \frac{x^3 y^2}{36} + \frac{x^4 y^2}{64} + \dots
\end{align*}
To fulfil the requirement, as discussed in previous examples, we consider the function 
\begin{align}
    f(x,y) = 1+ x+y +\mathrm{Li}_{2,2}(x-y, x+y)
\end{align}
and find the CA of $f(x,y)$ of order 10. Later it is massaged to yield the CA of $\mathrm{Li}_{2,2}(x, y) $. 

In the Table \eqref{table:tableLi22}, we compute the  CAs of the 
$\mathrm{Li}_{2,2}(x, y)$  series appearing in the RHS of Eq. \eqref{eqn:stuffle}, Eq. \eqref{eqn:inversion} and Eq. \eqref{eqn:stuffleinversion} (denoted as CA1, CA2 and CA3 respectively) and compare with the result of the \texttt{GINAC} \cite{Vollinga:2004sn} implementation through the \texttt{PolyLogTools} 
  package \cite{Duhr:2019tlz}. Here, CA0 denotes the CA of the defining series (i.e., Eq. \eqref{eqn:Li_def}).
It is to be noted that, the $\text{Li}_n$ series appearing in the expression of the analytic continuations of the $\mathrm{Li}_{2,2}(x, y)$  can be approximated using the Pad\'e approximation. However, those are evaluated with the inbuilt \mt commands.

We take values of $(|x|,|y|)$ from $(0.1,0.1)$ to $(2.1,2.1)$ in the interval of 0.5 for each of the variables and compute the CAs at these points. A small negative imaginary part is added to both the variables in order to avoid the branch cut issues of polylogarithms, which is not shown explicitly in Table \eqref{table:tableLi22}.  The first column of the Table indicates which region of convergences of the series $\mathrm{Li}_{2,2}(x, y)$ (Eq. \eqref{eq:rocLi}) the selected point belongs to. We observe that, in some instances, even if the chosen point does not belong to the region of convergence of the series, still its CA produces the correct result when compared to the \texttt{GINAC} implementaion.

\begin{table}[htbp]
\caption{Table of values of $\text{Li}_{2,2}(x,y)$}
\centering
\small
\begin{tabular}{|l|l|l|l|l|l|l|}
\hline
ROCs                                                                  & $\{|x|,|y|\}$ & CA0                                                                                       & CA1                                                                                       & CA2                                                              & CA3                                                                 & \texttt{GINAC} o/p                                                                          \\ \hline
\begin{tabular}[c]{@{}c@{}}\{True,True,\\ False,False\}\end{tabular}  & \{0.1,0.1\} & \begin{tabular}[c]{@{}c@{}}0.000262074 \\ $-7.99326\times 10^{-13}$ I\end{tabular} & \begin{tabular}[c]{@{}c@{}}0.000262074\\  $-7.99326\times10^{-13} $I\end{tabular} & \begin{tabular}[c]{@{}c@{}}7.01781\\  +21.689 I\end{tabular}     & \begin{tabular}[c]{@{}c@{}}-7.01729\\ -21.689 I\end{tabular}        & \begin{tabular}[c]{@{}c@{}}0.00026207\\ $+0. \times10^{-13}$ I\end{tabular} \\ \hline
\begin{tabular}[c]{@{}c@{}}\{True,True,\\ False,False\}\end{tabular}  & \{0.1,0.6\} & \begin{tabular}[c]{@{}c@{}}0.00158144 \\ $-3.51482\times10^{-12}$ I\end{tabular}  & \begin{tabular}[c]{@{}c@{}}0.00158144\\  $-3.51482\times10^{-12}$ I\end{tabular}  & \begin{tabular}[c]{@{}c@{}}14.0274\\  +2.56458 I\end{tabular}    & \begin{tabular}[c]{@{}c@{}}48.7329\\  +9.35645 I\end{tabular}       & \begin{tabular}[c]{@{}c@{}}0.0015814\\ $+0.\times10^{-12}$ I\end{tabular}  \\ \hline
\begin{tabular}[c]{@{}c@{}}\{True,False,\\ False,False\}\end{tabular} & \{0.1,1.1\} & \begin{tabular}[c]{@{}c@{}}0.00291623 \\ $-6.27839\times10^{-12}$ I\end{tabular}  & \begin{tabular}[c]{@{}c@{}}0.0341918 \\ -0.0307264 I\end{tabular}                         & \begin{tabular}[c]{@{}c@{}}9.65567 \\ -0.0954793 I\end{tabular}  & \begin{tabular}[c]{@{}c@{}}-1.70713\\ +3.84641 I\end{tabular}       & \begin{tabular}[c]{@{}c@{}}0.0029162\\ $+0.\times10^{-12}$ I\end{tabular}  \\ \hline
\begin{tabular}[c]{@{}c@{}}\{True,False,\\ False,False\}\end{tabular} & \{0.1,1.6\} & \begin{tabular}[c]{@{}c@{}}0.0042668\\  $-9.09157\times10^{-12}$ I\end{tabular}   & \begin{tabular}[c]{@{}c@{}}-4.73447\\ -0.151521 I\end{tabular}                            & \begin{tabular}[c]{@{}c@{}}7.29092 \\ -0.995102 I\end{tabular}   & \begin{tabular}[c]{@{}c@{}}-0.137509\\ +1.32557 I\end{tabular}      & \begin{tabular}[c]{@{}c@{}}0.0042671\\ $+0.\times10^{-12}$ I\end{tabular}  \\ \hline
\begin{tabular}[c]{@{}c@{}}\{True,False,\\ False,False\}\end{tabular} & \{0.1,2.1\} & \begin{tabular}[c]{@{}c@{}}0.005625\\  $-1.19236\times10^{-11}$ I\end{tabular}    & \begin{tabular}[c]{@{}c@{}}0.288087 \\ -0.239188 I\end{tabular}                           & \begin{tabular}[c]{@{}c@{}}5.78188\\  -1.35315 I\end{tabular}    & \begin{tabular}[c]{@{}c@{}}0.0935534 \\ +0.632078 I\end{tabular}    & \begin{tabular}[c]{@{}c@{}}0.0056347\\ $+0.\times10^{-11}$ I\end{tabular}  \\ \hline
\begin{tabular}[c]{@{}c@{}}\{True,True,\\ False,False\}\end{tabular}  & \{0.6,0.1\} & \begin{tabular}[c]{@{}c@{}}0.0128541 \\ $-1.82808\times10^{-11}$ I\end{tabular}   & \begin{tabular}[c]{@{}c@{}}0.0128541 \\ $-1.82808\times10^{-11}$ I\end{tabular}   & \begin{tabular}[c]{@{}c@{}}-48.7185\\ -9.35645 I\end{tabular}    & \begin{tabular}[c]{@{}c@{}}-14.013\\ -2.56458 I\end{tabular}        & \begin{tabular}[c]{@{}c@{}}0.0128541\\ $+0.\times10^{-11}$ I\end{tabular}  \\ \hline
\begin{tabular}[c]{@{}c@{}}\{True,True,\\ False,False\}\end{tabular}  & \{0.6,0.6\} & \begin{tabular}[c]{@{}c@{}}0.0803142 \\ $-4.80714\times10^{-11}$ I\end{tabular}   & \begin{tabular}[c]{@{}c@{}}0.0803142 \\ $-4.80714\times10^{-11}$ I\end{tabular}   & \begin{tabular}[c]{@{}c@{}}-4.81735\\ -3.62341 I\end{tabular}    & \begin{tabular}[c]{@{}c@{}}4.97798 \\ +3.62341 I\end{tabular}       & \begin{tabular}[c]{@{}c@{}}0.080314\\ $+0.\times10^{-11}$ I\end{tabular}   \\ \hline
\begin{tabular}[c]{@{}c@{}}\{True,False,\\ False,False\}\end{tabular} & \{0.6,1.1\} & \begin{tabular}[c]{@{}c@{}}0.154466 \\ $-8.29067\times10^{-11}$ I\end{tabular}    & \begin{tabular}[c]{@{}c@{}}0.298723\\  -0.217858 I\end{tabular}                           & \begin{tabular}[c]{@{}c@{}}-1.60591\\ -1.96513 I\end{tabular}    & \begin{tabular}[c]{@{}c@{}}-0.175139\\ +0.12099 I\end{tabular}      & \begin{tabular}[c]{@{}c@{}}0.15447\\ $+0.\times10^{-11}$ I\end{tabular}    \\ \hline
\begin{tabular}[c]{@{}c@{}}\{True,False,\\ False,False\}\end{tabular} & \{0.6,1.6\} & \begin{tabular}[c]{@{}c@{}}0.238586 \\ $-1.26853\times10^{-10}$ I\end{tabular}    & \begin{tabular}[c]{@{}c@{}}0.842343\\  -1.07432 I\end{tabular}                            & \begin{tabular}[c]{@{}c@{}}0.62006 \\ -1.22965 I\end{tabular}    & \begin{tabular}[c]{@{}c@{}}0.239959 \\ +0.0000562957 I\end{tabular} & \begin{tabular}[c]{@{}c@{}}0.23891\\ $+0.\times10^{-10}$ I\end{tabular}    \\ \hline
\begin{tabular}[c]{@{}c@{}}\{False,False,\\ False,True\}\end{tabular} & \{0.6,2.1\} & \begin{tabular}[c]{@{}c@{}}0.326298 \\ $-1.66348\times10^{-10}$ I\end{tabular}    & \begin{tabular}[c]{@{}c@{}}2.35114 \\ -1.68944 I\end{tabular}                             & \begin{tabular}[c]{@{}c@{}}1.62505 \\ -0.889454 I\end{tabular}   & \begin{tabular}[c]{@{}c@{}}0.344628 \\ -0.00749342 I\end{tabular}   & \begin{tabular}[c]{@{}c@{}}0.34449\\ -0.00749 I\end{tabular}                       \\ \hline
\begin{tabular}[c]{@{}c@{}}\{False,True,\\ False,False\}\end{tabular} & \{1.1,0.1\} & \begin{tabular}[c]{@{}c@{}}0.0563709 \\ $-1.65742\times10^{-10}$ I\end{tabular}   & \begin{tabular}[c]{@{}c@{}}0.0876465 \\ -0.0307264 I\end{tabular}                         & \begin{tabular}[c]{@{}c@{}}1.79769 \\ -3.87714 I\end{tabular}    & \begin{tabular}[c]{@{}c@{}}-9.56511\\ +0.0647529 I\end{tabular}     & \begin{tabular}[c]{@{}c@{}}0.087647\\ -0.030726 I\end{tabular}                     \\ \hline
\begin{tabular}[c]{@{}c@{}}\{False,True,\\ False,False\}\end{tabular} & \{1.1,0.6\} & \begin{tabular}[c]{@{}c@{}}0.436983 \\ $-7.06191\times10^{-10}$ I\end{tabular}    & \begin{tabular}[c]{@{}c@{}}0.58124 \\ -0.217858 I\end{tabular}                            & \begin{tabular}[c]{@{}c@{}}0.910845 \\ -0.338848 I\end{tabular}  & \begin{tabular}[c]{@{}c@{}}2.34162 \\ +1.74727 I\end{tabular}       & \begin{tabular}[c]{@{}c@{}}0.58124\\ -0.21786 I\end{tabular}                       \\ \hline
\begin{tabular}[c]{@{}c@{}}\{False,False,\\ True,False\}\end{tabular} & \{1.1,1.1\} & \begin{tabular}[c]{@{}c@{}}0.863226 \\ $-1.60629\times10^{-9}$ I\end{tabular}     & \begin{tabular}[c]{@{}c@{}}1.55121 \\ -1.17132 I\end{tabular}                             & \begin{tabular}[c]{@{}c@{}}1.20717 \\ -0.58566 I\end{tabular}    & \begin{tabular}[c]{@{}c@{}}1.20727 \\ -0.58566 I\end{tabular}       & \begin{tabular}[c]{@{}c@{}}1.20722\\ -0.58566 I\end{tabular}                       \\ \hline
\begin{tabular}[c]{@{}c@{}}\{False,False,\\ True,False\}\end{tabular} & \{1.1,1.6\} & \begin{tabular}[c]{@{}c@{}}1.06793 \\ $-1.38643\times10^{-9}$ I\end{tabular}      & \begin{tabular}[c]{@{}c@{}}17.5197\\  -3.52497 I\end{tabular}                             & \begin{tabular}[c]{@{}c@{}}1.66013 \\ -1.23401 I\end{tabular}    & \begin{tabular}[c]{@{}c@{}}1.66016 \\ -1.23401 I\end{tabular}       & \begin{tabular}[c]{@{}c@{}}1.6602\\ -1.2340 I\end{tabular}                         \\ \hline
\begin{tabular}[c]{@{}c@{}}\{False,False,\\ True,False\}\end{tabular} & \{1.1,2.1\} & \begin{tabular}[c]{@{}c@{}}2.75708 \\ $-2.17691\times10^{-9}$ I\end{tabular}      & \begin{tabular}[c]{@{}c@{}}13.6042\\  -5.00396 I\end{tabular}                             & \begin{tabular}[c]{@{}c@{}}1.89607\\ -1.899 I\end{tabular}       & \begin{tabular}[c]{@{}c@{}}1.89609 \\ -1.899 I\end{tabular}         & \begin{tabular}[c]{@{}c@{}}1.8961\\ -1.8990 I\end{tabular}                         \\ \hline
\begin{tabular}[c]{@{}c@{}}\{False,True,\\ False,False\}\end{tabular} & \{1.6,0.1\} & \begin{tabular}[c]{@{}c@{}}4.82045 \\ $+3.02096\times10^{-8}$ I\end{tabular}      & \begin{tabular}[c]{@{}c@{}}0.0817101 \\ -0.151521 I\end{tabular}                          & \begin{tabular}[c]{@{}c@{}}0.223486\\  -1.47709 I\end{tabular}   & \begin{tabular}[c]{@{}c@{}}-7.20494\\ +0.843581 I\end{tabular}      & \begin{tabular}[c]{@{}c@{}}0.08171\\ -0.15152 I\end{tabular}                       \\ \hline
\begin{tabular}[c]{@{}c@{}}\{False,True,\\ False,False\}\end{tabular} & \{1.6,0.6\} & \begin{tabular}[c]{@{}c@{}}-0.121148\\ $-1.37615\times10^{-9}$ I\end{tabular}     & \begin{tabular}[c]{@{}c@{}}0.482609 \\ -1.07432 I\end{tabular}                            & \begin{tabular}[c]{@{}c@{}}0.481236\\  -1.07438 I\end{tabular}   & \begin{tabular}[c]{@{}c@{}}0.101135 \\ +0.155323 I\end{tabular}     & \begin{tabular}[c]{@{}c@{}}0.48228\\ -1.07432 I\end{tabular}                       \\ \hline
\begin{tabular}[c]{@{}c@{}}\{False,False,\\ True,False\}\end{tabular} & \{1.6,1.1\} & \begin{tabular}[c]{@{}c@{}}-15.3222\\ $-6.56091\times10^{-8}$ I\end{tabular}      & \begin{tabular}[c]{@{}c@{}}1.1296 \\ -3.52497 I\end{tabular}                              & \begin{tabular}[c]{@{}c@{}}0.537364 \\ -2.29096 I\end{tabular}   & \begin{tabular}[c]{@{}c@{}}0.537392 \\ -2.29096 I\end{tabular}      & \begin{tabular}[c]{@{}c@{}}0.5374\\ -2.2910 I\end{tabular}                         \\ \hline
\begin{tabular}[c]{@{}c@{}}\{False,False,\\ True,False\}\end{tabular} & \{1.6,1.6\} & \begin{tabular}[c]{@{}c@{}}-12.7656\\ $-3.05954\times10^{-8}$ I\end{tabular}      & \begin{tabular}[c]{@{}c@{}}13.224\\  -6.69136 I\end{tabular}                              & \begin{tabular}[c]{@{}c@{}}0.229189 \\ -3.34568 I\end{tabular}   & \begin{tabular}[c]{@{}c@{}}0.229189 \\ -3.34568 I\end{tabular}      & \begin{tabular}[c]{@{}c@{}}0.2292\\ -3.3457 I\end{tabular}                         \\ \hline
\begin{tabular}[c]{@{}c@{}}\{False,False,\\ True,False\}\end{tabular} & \{1.6,2.1\} & \begin{tabular}[c]{@{}c@{}}-0.742427\\ $-1.23116\times10^{-9}$ I\end{tabular}     & \begin{tabular}[c]{@{}c@{}}-4.31387\\ -8.33243 I\end{tabular}                             & \begin{tabular}[c]{@{}c@{}}-0.203749\\ -4.20211 I\end{tabular}   & \begin{tabular}[c]{@{}c@{}}-0.203749\\ -4.20211 I\end{tabular}      & \begin{tabular}[c]{@{}c@{}}-0.2037\\ -4.2021 I\end{tabular}                        \\ \hline
\begin{tabular}[c]{@{}c@{}}\{False,True,\\ False,False\}\end{tabular} & \{2.1,0.1\} & \begin{tabular}[c]{@{}c@{}}-0.248007\\ $+2.0216\times10^{-10}$ I\end{tabular}     & \begin{tabular}[c]{@{}c@{}}0.0344549 \\ -0.239188 I\end{tabular}                          & \begin{tabular}[c]{@{}c@{}}-0.0534735\\ -0.871266 I\end{tabular} & \begin{tabular}[c]{@{}c@{}}-5.7418\\ +1.11396 I\end{tabular}        & \begin{tabular}[c]{@{}c@{}}0.03445\\ -0.23919 I\end{tabular}                       \\ \hline
\begin{tabular}[c]{@{}c@{}}\{False,False,\\ True,False\}\end{tabular} & \{2.1,0.6\} & \begin{tabular}[c]{@{}c@{}}-1.96839\\ $-1.30701\times10^{-10}$ I\end{tabular}     & \begin{tabular}[c]{@{}c@{}}0.0564561 \\ -1.68944 I\end{tabular}                           & \begin{tabular}[c]{@{}c@{}}0.0381255 \\ -1.68195 I\end{tabular}  & \begin{tabular}[c]{@{}c@{}}-1.24229\\ -0.799988 I\end{tabular}      & \begin{tabular}[c]{@{}c@{}}0.0383\\ -1.6819 I\end{tabular}                         \\ \hline
\begin{tabular}[c]{@{}c@{}}\{False,False,\\ True,False\}\end{tabular} & \{2.1,1.1\} & \begin{tabular}[c]{@{}c@{}}-12.3171\\ $-1.0927\times10^{-8}$ I\end{tabular}       & \begin{tabular}[c]{@{}c@{}}-1.47002\\ -5.00396 I\end{tabular}                             & \begin{tabular}[c]{@{}c@{}}-0.609029\\ -3.10496 I\end{tabular}   & \begin{tabular}[c]{@{}c@{}}-0.609011\\ -3.10496 I\end{tabular}      & \begin{tabular}[c]{@{}c@{}}-0.6090\\ -3.1050 I\end{tabular}                        \\ \hline
\begin{tabular}[c]{@{}c@{}}\{False,False,\\ True,False\}\end{tabular} & \{2.1,1.6\} & \begin{tabular}[c]{@{}c@{}}2.63339\\  $-2.38178\times10^{-9}$ I\end{tabular}      & \begin{tabular}[c]{@{}c@{}}-0.938048\\ -8.33243 I\end{tabular}                            & \begin{tabular}[c]{@{}c@{}}-1.47673\\ -4.13032 I\end{tabular}    & \begin{tabular}[c]{@{}c@{}}-1.47673\\ -4.13032 I\end{tabular}       & \begin{tabular}[c]{@{}c@{}}-1.4767\\ -4.1303 I\end{tabular}                        \\ \hline
\begin{tabular}[c]{@{}c@{}}\{False,False,\\ True,False\}\end{tabular} & \{2.1,2.1\} & \begin{tabular}[c]{@{}c@{}}3.57931\\  $-3.54955\times10^{-9}$ I\end{tabular}      & \begin{tabular}[c]{@{}c@{}}-8.29019\\ -9.78067 I\end{tabular}                             & \begin{tabular}[c]{@{}c@{}}-2.35544\\ -4.89033 I\end{tabular}    & \begin{tabular}[c]{@{}c@{}}-2.35544\\ -4.89033 I\end{tabular}       & \begin{tabular}[c]{@{}c@{}}-2.3554\\ -4.8903 I\end{tabular}                        \\ \hline
\end{tabular}
\label{table:tableLi22}
\end{table}

The defining series of $\mathrm{Li}_{2,2}(x, y)$ (i.e., Eq. \eqref{eqn:Li_def}) is very slowly convergent around the point $\left(|x|, |y|\right) = (1,1)$, where the presented acceleration technique is found to be useful. In the Table \ref{table:Li22at11}, we show a comparison of the values obtained from summing the series defined in Eq. \eqref{eqn:Li_def} and its CA with order $(o)$ varying from 5 to 20. In the right most column, the number of terms used for the summation is indicated, which we choose for convenience to be $(2o+1)^2$ for a given order $o$. We clearly see that the value obtained using CA is more accurate compared to the value obtained by summing the series. Note that
\begin{align}
    \mathrm{Li}_{2,2}(1, 1) = \frac{\pi^4}{120} = 0.811742425283354
\end{align}

Higher order CA is needed to obtain more accurate result.

\begin{table}[H]
\caption{Table of values of  
$\mathrm{Li}_{2,2}(x, y)$ at $(1,1)$}
\centering
\label{table:Li22at11}
\begin{tabular}{|c|c|c|c|}
\hline
Order  & value from CA     & value of $\mathrm{Li}_{2,2}(x, y)$  & number of terms \\ 
  &  &  from series & in the summation \\\hline
5     & 0.726068215009552 & 0.690568727620971       & 121             \\ \hline
6     & 0.743812703465901 & 0.706590246937065       & 169             \\ \hline
7     & 0.763247794268115 & 0.718828257083165       & 225             \\ \hline
8     & 0.771956662975054 & 0.728488465125224       & 289             \\ \hline
9     & 0.780513212719172 & 0.736311803998949       & 361             \\ \hline
10    & 0.785440863070842 & 0.742779189825413       & 441             \\ \hline
11    & 0.789944433469386 & 0.748216626989709       & 529             \\ \hline
12    & 0.793057205586444 & 0.752853064361067       & 625             \\ \hline
13    & 0.795665279868944 & 0.756854084618387       & 729             \\ \hline
14    & 0.797835077625447 & 0.760342450837344       & 841             \\ \hline
15    & 0.799406544586245 & 0.763411133663296       & 961             \\ \hline
16    & 0.801148912443423 & 0.766131848591896       & 1089            \\ \hline
17    & 0.802067561873525 & 0.768560812629968       & 1225            \\ \hline
18    & 0.806191214220949 & 0.770742723657730       & 1369            \\ \hline
19    & 0.803711956950532 & 0.772713572001726       & 1521            \\ \hline
20    & 0.804066077181726 & 0.774502665799193       & 1681            \\ \hline
\end{tabular}
\end{table}

\section{Summary and discussion}\label{sec:sumdiss}
We present an automation to evaluate Chisholm approximants \cite{chisholm1973rational} for two variable series, in \mt. The CA are a natural generalization of the well-known \pd approximants for the one variable case. They have the advantage of reducing to the latter when one of the variables in the approximants is set to 0. They also have various other symmetric and group properties. For the moment, we just focus on the diagonal approximants. We present several examples to demonstrate the usage of the package using some elementary functions such as $\exp(\frac{x+y}{2}), \sin(\frac{x+y}{2}), \sinh(\frac{x+y}{2})$ and $\log(1+x+y)$. For the case of $\log(1+x+y)$ we see that the CA is also valid where the Taylor series, using which the approximants have been constructed from is not valid. This shows that the CA also performs the analytic continuation in some cases. Furthermore, we show some applications of these approximants in physics. We consider examples of hypergeometric functions such as Appell $F_1$, $F_2$ and $\mathrm{Li}_{2,2}(x,y)$ show the utility of these approximants for their evaluation purposes. As has been shown for the case of $\mathrm{Li}_{2,2}(x,y)$ the 
CA can also be used to accelerate the convergence of the double series. We further present the application of these approximants in the study of critical phenomena in condensed matter physics. 

We emphasise that the method presented in this paper is not the only way to obtain the two-variable approximant. Other methods presented in \cite{jones1974generation,cuyt1999well,guillaume1997nested,guillaume1998convergence} can also be used to find the two-variable approximants. As a future problem, it would be interesting to study and compare the efficiency of these methods and the results obtained using them. We also notice that the present implementation of CA is symmetric and one can also look for possibilities of CA which break the symmetry in one of the other ways as has been already mentioned in \cite{chisholm1973rational}. A simple way to break the symmetry of the CA obtained is to consider the simple off-diagonal Chisholm approximants \cite{graves1974calculation} analogous to \pd approximant case. Another exciting direction of study would be to develop a package for $N-$variable approximant\cite{chisholm1974rational}, using the simple generalization of Chisholm's method for the two-variable case. Similar to the two variable case there are various methods to form the rational approximants for the $N-$variables \cite{jones1976general,saff1977pade,levin1976general,baker1961pade,cuyt1987recursive,cuyt1996nuttall} case which can be studied and compared. 

\section{Acknowledgement}
We would like to thank B. Ananthanarayan for the useful suggestions and comments on the manuscript. We would also like to thank Sudeepan Datta for stimulating discussions. This work is part of TP's doctoral thesis.

\bibliographystyle{unsrt}
\bibliography{reference}
\end{document}